# EventLFM: Event Camera integrated Fourier Light Field Microscopy for Ultrafast 3D imaging

Ruipeng Guo[1], Qianwan Yang[1], Andrew S. Chang[4], Guorong Hu[1], Joseph Greene[1], Christopher V. Gabel[3,4], Sixian You[5], and Lei Tian[1,2,3,*]

[1]Department of Electrical and Computer Engineering, Boston University, Boston, MA 02215, USA.
[2]Department of Biomedical Engineering, Boston University, Boston, MA 02215, USA.
[3]Neurophotonics Center, Boston University, Boston, MA 02215, USA.
[4]Department of Physiology and Biophysics, Boston University, Boston, MA 02215, USA.
[5]Research Laboratory of Electronics (RLE) in the Department of Electrical Science and Engineering, Massachusetts Institute of Technology, Cambridge, MA 02139, USA.
*Correspondence: leitian@bu.edu, Tel.: 1-617-353-1334

Authors' email:

Ruipeng Guo: rguo@bu.edu

Qianwan Yang: yaw@bu.edu

Andrew S. Chang: aschang@bu.edu

Guorong Hu: grhu@bu.edu

Joseph Greene: joeg18@bu.edu

Christopher V. Gabel: cvgabel@bu.edu

Sixian You: sixian@mit.edu

Lei Tian: leitian@bu.edu

**Abstract**

Ultrafast 3D imaging is indispensable for visualizing complex and dynamic biological processes. Conventional scanning-based techniques necessitate an inherent trade-off between acquisition speed and space-bandwidth product (SBP). Emerging single-shot 3D wide-field techniques offer a promising alternative but are bottlenecked by the synchronous readout constraints of conventional CMOS systems, thus restricting data throughput to maintain high SBP at limited frame rates. To address this, we introduce EventLFM, a straightforward and cost-effective system that overcomes these challenges by integrating an event camera with Fourier light field microscopy (LFM), a state-of-the-art single-shot 3D wide-field imaging technique. The event camera operates on a novel asynchronous readout architecture, thereby bypassing the frame rate limitations inherent to conventional CMOS systems. We further develop a simple and robust event-driven LFM reconstruction algorithm that can reliably reconstruct 3D dynamics from the unique spatiotemporal measurements captured by EventLFM. Experimental results demonstrate that EventLFM can robustly reconstruct fast-moving and rapidly blinking 3D fluorescent samples at kHz frame rates. Furthermore, we highlight EventLFM's capability for imaging of blinking neuronal signals in scattering mouse brain tissues and 3D tracking of GFP-labeled neurons in freely moving *C. elegans*. We believe that the combined ultrafast speed and large 3D SBP offered by EventLFM may open up new possibilities across many biomedical applications.

# 1 Introduction

High-speed volumetric imaging is an indispensable tool for investigating dynamic biological processes [1]. Traditional scanning-based 3D imaging techniques such as confocal microscopy [2], two-photon microscopy [3] and light-sheet microscopy [4] offer high spatial resolution. However, their data acquisition speeds are often constrained by the need for beam scanning. Consequently, these techniques suffer from an inherent trade-off between temporal resolution and the space-bandwidth product (SBP), measured by the ratio of the 3D field-of-view (FOV) to the spatial resolution.

Single-shot 3D widefield imaging techniques circumvent this trade-off by computational imaging. These methods first encode 3D information into 2D multiplexed measurements and then reconstruct the 3D volume computationally. Examples of such techniques include light field microscopy (LFM) [5-9], lensless imaging [10-12], and point-spread-function (PSF) engineering [13, 14]. Conventional LFM works by inserting a microlens array (MLA) into the native image plane of a widefield microscope [5]. Each microlens captures unique spatial and angular information from a sample, allowing for subsequent computational 3D reconstructions without scanning. However, this configuration suffers from intrinsic limitations, such as inconsistent spatial resolution due to uneven spatial sampling across the MLA. Recently, Fourier LFM has emerged as a solution to alleviate the limitations of the conventional LFM by positioning the MLA at the Fourier or pupil plane of a microscope [6, 7]. By recording the light field information at the Fourier domain, Fourier LFM ensures uniform angular sampling, which allows the technique to achieve a consistently high spatial resolution across the recovered volume. Despite these advancements, the *synchronous* readout constraints of traditional CMOS cameras remain a fundamental bottleneck for single-shot 3D wide-field techniques. Although it is possible to increase the frame rate by restricting the readout to only a specific region of interest (ROI) from the CMOS sensor, this inevitably results in a reduced SBP. As a result, existing LFM techniques typically operated below 100 Hz at the full-frame resolution. This constraint hinders their applications in capturing ultrafast dynamic biological processes that may exceed kilohertz (kHz), such as voltage signals in mammalian brains [15], blood flow dynamics [16] and muscle tissue contraction [17], thus leaving a significant technological gap yet to be bridged.

To address these technological limitations, ultrafast imaging strategies have emerged, showing promise in various applications, such as characterization of ultrafast optical phenomena [18, 19], fluorescence lifetime imaging [20, 21], non-line-of-sight imaging [22], voltage imaging in mouse brains [23-25], and neurovascular dynamics recording [26]. Despite their merits, these techniques often come with their own trade-offs, such as the need for high-power illumination, which can be phototoxic to live biological specimens, and the need for expensive, specialized, and complicated optical systems.

Recently, event cameras have garnered significant attention over the past decade for their capability to provide kHz or higher frame rates while offering flexible integration into various platforms [27]. Unlike traditional CMOS cameras that record information from the full frame synchronously and at set time intervals, the event camera employs an *asynchronous* readout architecture. Each pixel on the event camera independently generates "event" readouts based on the changes in the pixel-level brightness over time. Each event recording contains information about the pixel's position, timestamp, and polarity of the brightness change, allowing for ultra-high temporal resolution and reduced latency [28]. As a result, event camera enables recording of ultrafast signal changes at speeds exceeding 10 kHz limited only by pixel latency. In addition, event cameras are implicitly sensitive to changes in the logarithm of the photocurrent, allowing them to achieve high dynamic range that exceeds 120dB [29]. Leveraging these unique attributes, event cameras have shown promise across diverse applications, ranging from self-driving [30] and gesture recognition [31] to single-molecule localization microscopy [32, 33].

In this work, we introduce EventLFM, a novel ultrafast, single-shot 3D imaging technique that integrates an event camera into a Fourier LFM system, as illustrated in Fig. 1(a). We develop a simple event-driven LFM reconstruction algorithm to reliably reconstruct 3D dynamics from EventLFM's spatiotemporal measurements. We experimentally demonstrate the applicability of EventLFM for 3D fluorescence imaging on various samples, achieving speeds of up to 1 kHz, effectively bridging the existing technological gaps in capturing ultrafast dynamic 3D processes.

To elucidate the method, Fig. 1(a) shows an example involving a suspension of fluorescent beads that traverse various trajectories within a 3D space. EventLFM captures a stream of events, as depicted in Fig. 1(b), which arise from instantaneous changes in intensity due to the rapid displacement of these beads across the FOV. Our event-driven LFM reconstruction algorithm works by first converting the acquired events into "conventional" 2D frame-based representations. This conversion is performed through a time-surface based method that leverages both spatial and temporal correlations among events over a predefined temporal-spatial window [34, 35]. The algorithm assigns values to each pixel based on accumulated historical data, which is shown in Fig. 1(c). Subsequently, these generated time-surface maps are cropped into 5 x 5 views and undergo a 3D reconstruction via the light field refocusing algorithm [36], as illustrated in Fig. 1(d). This refocused volume can also be enhanced by a deep learning module to improve the 3D resolution and suppress the noise artifacts, as illustrated in Fig. 1(f). A representative frame depicting the 3D reconstruction of the fluorescent beads is provided in Fig. 1(e) displaying a depth color-coded map. Finally, to encapsulate the entire 4D information, a spatiotemporal reconstruction is visualized. This entails performing EventLFM reconstruction from the event measurements with an equivalent 1 kHz frame rate, spanned across a 45 ms time window. Fig. 1(g) shows the recovered 3D trajectories of the fast-moving fluorescent beads with a depth color-coded map encoding the temporal information.

We provide a quantitative evaluation of EventLFM's 4D imaging capabilities across a range of fast dynamic samples. This includes fast-moving fluorescent beads subjected to both controlled and random 3D motions, as well as rapid blinking beads that operate at frequencies up to 1 kHz, both with and without 3D motions. In addition, we showcase a demonstrative experiment on the imaging of rapidly blinking neuronal signals simulated through either uniform or targeted illumination in a 75 µm thick mouse brain section, as illustrated in Fig. 1(h), and demonstrate EventLFM's ability to significantly enhance signal contrast in scattering tissues by rejecting the temporally slowly varying background in the raw event measurements. Furthermore, we demonstrate 3D tracking of labeled neurons in multiple freely moving *C. elegans*. Our results collectively demonstrate the robust and ultrafast 3D imaging capabilities of EventLFM, thereby underscoring its potential for elucidating intricate 3D dynamical phenomena within biological systems.

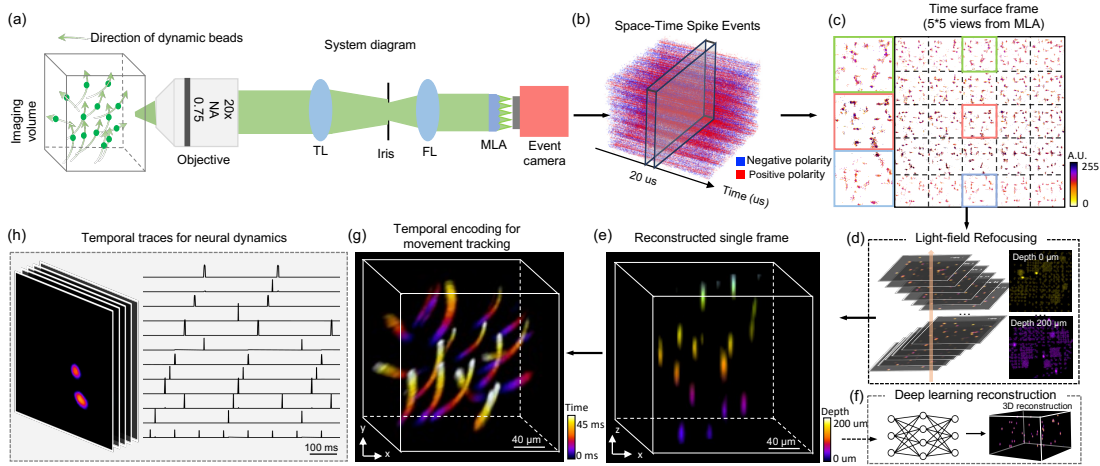

**Figure 1: Overview of EventLFM.** (a) Schematics of the setup. (b) Space-time event spike stream captured by the event camera, where blue and red indicate negative and positive intensity changes at individual pixels, respectively. (c) Frames are generated from the raw event stream using the time surface algorithm within a 1 ms accumulation time interval. (d) Each time-surface frame is reconstructed using a light-field refocusing algorithm with an optional deep learning enhancement step. (e) Depth color-coded 3D light field reconstruction of the object. (f) An optional convolutional neural network (CNN) refines the refocused volume, enhancing the 3D resolution and suppressing

the noise artifacts. (g) Time color-coded 3D motion trajectory reconstructed over a 45 ms time span. (h) Illustrative experiment results on rapidly blinking neurons in a 75 μm thick, weakly scattering brain section. Temporal traces capturing dynamic neuronal activities are extracted from reconstructed frames with an effective frame rate of 1 kHz.

## 2  Results

### 2.1  System characterization

To validate the fidelity of the result of EventLFM, we conduct a comparative study with a standard Fourier LFM equipped with an sCMOS camera. First, we calibrate the 3D PSF for both systems. Given that the event camera can only capture dynamic objects, EventLFM PSFs are obtained from an event stream generated by a bead translating continuously through the system's depth of field (DOF) at 0.2 mm/s, as illustrated in Fig. 2(a). For the standard Fourier LFM, PSFs are obtained by scanning a 1 μm bead along the $z$-axis. Subsequently, we analyze the lateral and axial resolutions by computing the 3D modulation transfer function (MTF) for both systems, defined by the 3D Fourier spectrum of the calibrated 3D PSF [8]. Fig. 2(b) shows a strong agreement in both the 3D PSFs and the 3D MTFs between EventLFM and the standard Fourier LFM, thereby validating EventLFM's ability to achieve high spatial resolution at a markedly improved frame rate (1000 Hz vs. 30 Hz). We note that the PSF measurements from the event camera are noisier than that from the sCMOS camera due to more pronounced noise from the measured event stream. Additional performance metrics of standard Fourier LFM, such as FOV, DOF, and resolution, are elaborated in Section 1 of Supplement 1.

For an intuitive, side-by-side comparison, we simultaneously acquire data from a slowly moving 3D fluorescent beads phantom using both systems. Both datasets - comprising time-surface frames from EventLFM and sCMOS frames - are processed via the same light field refocusing algorithm to generate 3D reconstructions. Fig. 2(c) shows single-frame depth color-coded 3D reconstructions from both systems. The consistency between the two methods verifies EventLFM's fidelity in reconstructing depth information throughout the DOF. To further confirm that EventLFM provides consistent axial resolution, intensity profiles extracted from the same bead along the white dashed lines are compared in Fig. 2(d). For further validation, conventional widefield fluorescence microscopy (Plan Apo, 20X, 0.75 NA, Nikon) is also employed to capture a $z$-stack of the same phantom (see details in Section 1 of Supplement 1). A comparison of depth color-coded maximum intensity projections (MIPs) across all three methods is shown in Fig. 2(e). The results confirm EventLFM's capability for accurate volumetric reconstruction across the entire FOV. Intriguingly, we observe that the axial elongation achieved by EventLFM is slightly shorter than that achieved by the standard Fourier LFM, as evidenced in both the 3D reconstructions (Fig. 2(c)) and axial profiles of individual beads (Fig. 2(d)). We attribute this observation to the unique event-driven signal acquisition mechanism of the event camera. Specifically, an accumulation time of 1 ms necessitates sufficient power to trigger events. When the illumination power is low, only the central region of the beads has adequate intensity to generate such events, which in turn reduces the axial elongation in the reconstructions.

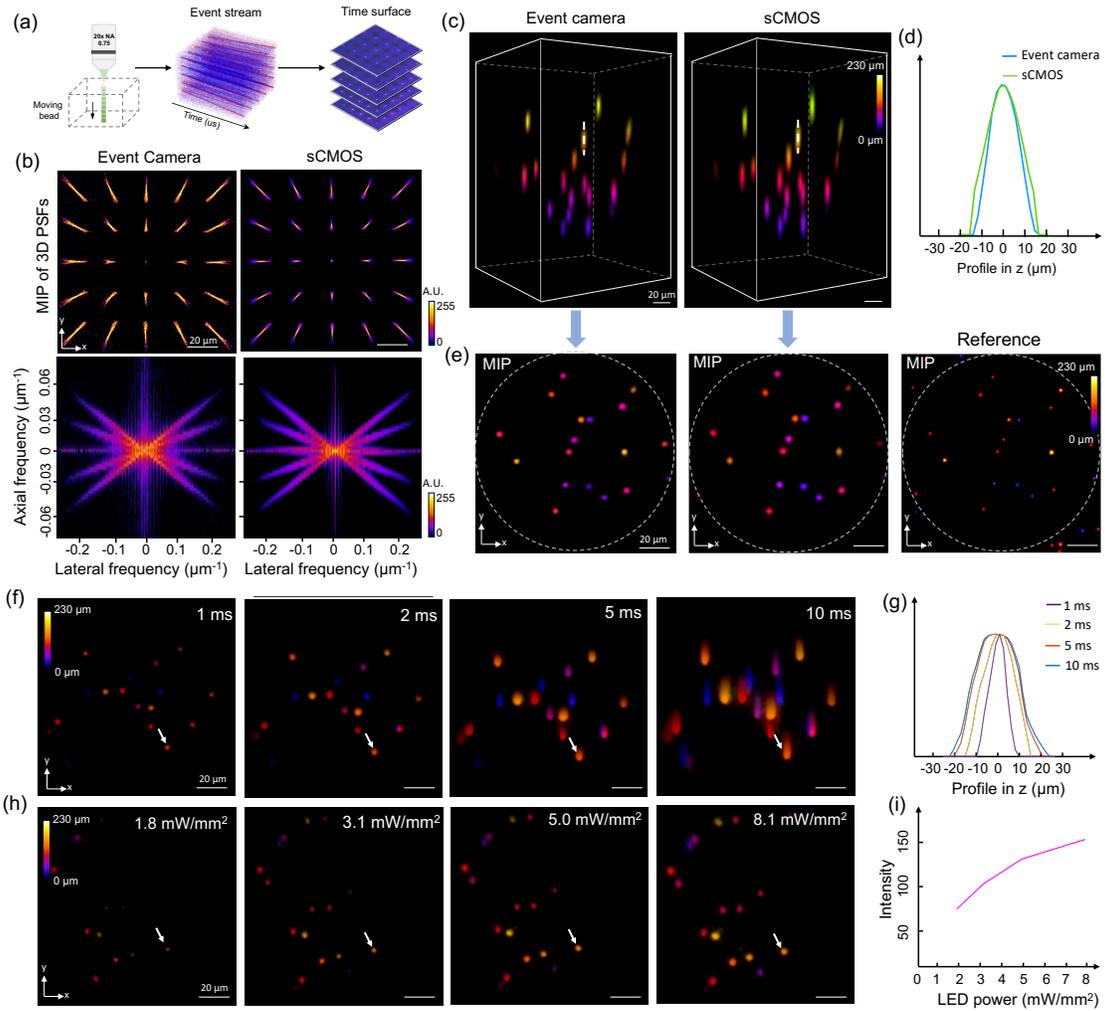

**Figure 2: Characterization of the EventLFM system.** (a) 3D PSF acquisition pipeline. The event camera captures the vertical motion of a single bead within the DOF. This raw event stream is subsequently converted into discrete frames using the time-surface algorithm. (b) Comparisons of MIPs of the 3D PSFs and 3D MTFs for both the EventLFM and a benchmark Fourier LFM using an sCMOS camera. (c) 3D reconstruction comparison of fluorescent beads distributed volumetrically, obtained using EventLFM and the benchmark Fourier LFM. (d) Axial resolution assessment based on profiles of a single bead from 3D reconstructions obtained with both the event camera (in blue) and the sCMOS camera (in green), demonstrating that both systems offer comparable axial resolution. (e) MIPs of the reconstructed 3D volumes from both LFM methods. For additional validation, a reference MIP is obtained through axial scanning of the same object using conventional widefield fluorescence microscopy. (f) Accumulation time sensitivity is analyzed by plotting the MIPs of EventLFM reconstructions at different time intervals. (g) Effect of accumulation time on axial elongation is assessed by plotting the axial intensity profiles of a bead as marked in (f), revealing an increase in axial elongation correlated with longer accumulation times. (h) Illumination power analysis is performed by visualizing the MIPs of EventLFM reconstructions for the same object but under varying LED power levels. (i) Correlation between the mean intensity of a reconstructed bead (as indicated in (h)) and the LED power levels.

We also characterize how EventLFM's performance is affected by key experimental parameters, specifically the accumulation time of the event camera and the illumination power. The raw event stream from the event camera exhibits a temporal resolution of 1 μs. When this data is transformed into frames, the user-defined accumulation time significantly influences the quality of the reconstruction. To demonstrate this, we image a fluorescent beads phantom moving at 2.5 mm/s along the *y* direction. Similar to conventional cameras, an elongated accumulation time leads to increased averaged intensity and enlarged/blurred bead images, as shown in Fig. 2(f) and axial profiles in Fig. 2(g). By properly selecting an appropriate accumulation time based on sample's brightness levels and event dynamics, the event camera can achieve superior resolution (See more discussion about setting the camera in Section 4 of Supplement 1). It should be noted that this parameter is adjusted in the post-processing step *without* impacting the data capture speed. Next, while the event stream inherently lacks information on absolute intensity, we observe its sensitivity to variations in object brightness levels, as shown in Fig. 2(h). Intuitively, this is because a larger intensity variation produces more events in quick succession. To demonstrate this, we record the same fluorescent beads phantom moving at 1 mm/s along the *y* direction under varying illumination powers, spanning 1.8 mW/mm$^2$ to 8.1 mW/mm$^2$, while keeping the accumulation time constant. The subsequent EventLFM reconstructions reveal a positive correlation between reconstructed intensity and illumination power, as depicted in Fig. 2(i).

## 2.2 Imaging of fast-moving objects

We substantiate the capability of EventLFM to reconstruct high-speed 3D motion, demonstrating its utility in capturing dynamical phenomena in biological contexts. First, we employ a motorized stage with a velocity range of 0.001 mm/s to 2.7 mm/s to execute controlled motion experiments. We image a 3D phantom comprising 2 μm fluorescent beads moving at 2.5 mm/s. EventLFM successfully reconstructs the rapid movements across all depths at an effective frame rate of 1 kHz, as illustrated in Fig. 3(a). Furthermore, to evaluate the imaging performance of EventLFM across various velocities, we extract an ROI and present consecutive frames in Fig. 3(b-c). Fig. 3(b) shows the accelerating process from 0.1 mm/s to 2.5 mm/s with acceleration of 2.5 mm/s$^2$. The trajectory's slope (dashed white curve) steepens over time, reflecting the increased speed. Fig. 3(c) provides consecutive frames within the same ROI at the peak velocity of 2.5 mm/s. Given the object's fixed velocity along the *x*-axis, the bead positions calculated from the motorized stage setting align well with the EventLFM reconstructions. In contrast, the standard Fourier LFM using the sCMOS camera operating at 30 Hz suffers from severe motion blur artifacts. We image the same object using the benchmark Fourier LFM system under static and slow-motion conditions (see details in Section 5 of Supplement 1).

Additionally, the system's lateral and axial resolution are quantitatively evaluated at various speeds by calculating the FWHM of a reconstructed single bead, as depicted in Fig. 3(d). The 3D resolutions are consistent across a velocity spectrum from 0.1 mm/s to 2.5 mm/s, demonstrating the EventLFM system's capability to reliably image objects in motion at differing velocities without resolution degradation, which further corroborate the robustness of our EventLFM system. These controlled experiments confirm EventLFM's efficacy in capturing rapid 3D dynamics at frame rates up to 1 kHz.

In the context of biological applications, the motion of many samples occurs over a gamut of velocities, directions, and depths. Acknowledging this complexity, we extend our EventLFM evaluations to scenarios involving uncontrolled complex 3D motion. Specifically, fluorescent beads are suspended in an alcohol-water droplet subjected to ultrasonic disintegration, inducing variable motion directions and velocities exceeding 2.5 mm/s. After performing EventLFM reconstructions, we present a depth color-coded MIP in Fig. 3(e). A selected sub-region, marked by a white dashed square, is subjected to volumetric rendering for 5 representative frames in Fig. 3(f). Leveraging the millisecond-level temporal resolution, we trace intricate trajectories (depicted as dotted lines with arrows) for individual beads. Notably, complex motion patterns - including depth fluctuations - are faithfully captured. For instance, a bead represented in blue in the frame labeled #00 in Fig. 3(f) exhibits helical movement through the volume over several microseconds. This affirms EventLFM's utility in characterizing complex, high-speed 3D dynamics.

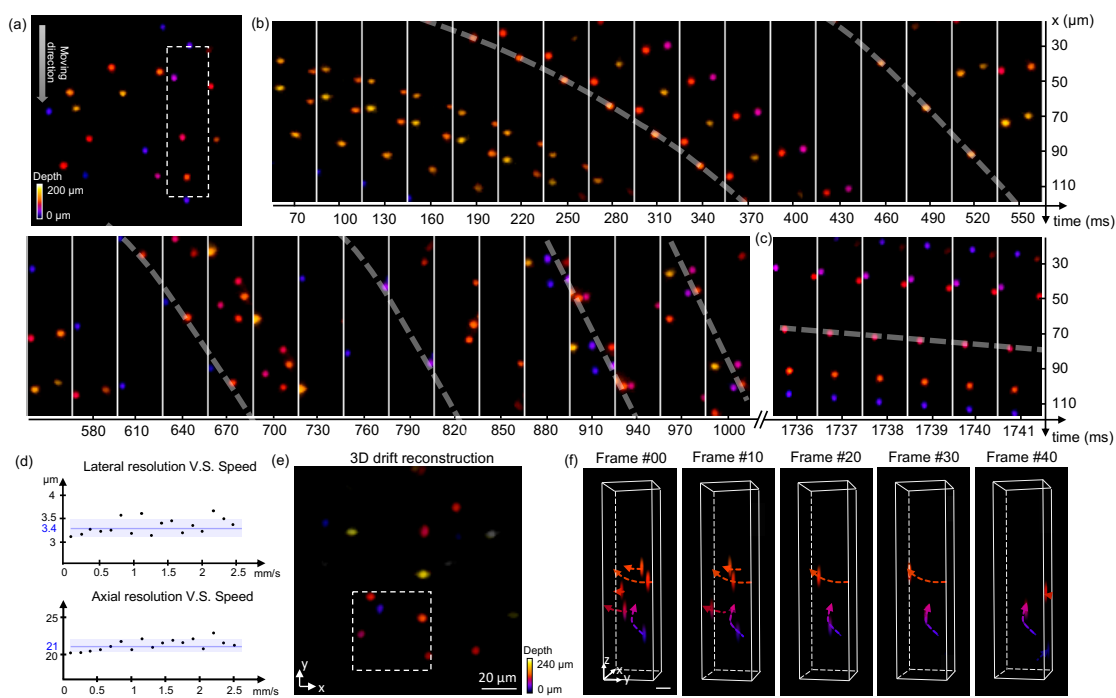

**Figure 3: EventLFM imaging of fast-moving objects.** (a-d) Results from objects with directional movement. (a) Depth color-coded MIP of a single reconstructed frame capturing a phantom comprising fluorescent beads. The beads move in a single direction, as denoted by the arrow, at a calibrated speed. (b) Zoom-in MIPs from 32 consecutive frames, revealing the unidirectional, continuous accelerating motion of the beads over time. The shadowed traces on the projections highlight the beads' motion trajectory over time, with an increasing slope indicative of accelerated movement. (c) Zoom-in MIPs from 6 consecutive frames, revealing the unidirectional, uniform motion of the beads on a millisecond scale at 2.5mm/s over time. (d) Lateral and axial resolutions across varying velocities are accessed through the FWHM of a single bead in the reconstruction at different frames. The 3D resolutions remain stable across speeds ranging from 0.1 mm/s to 2.5 mm/s, with the blue line indicating the average resolution, and the shaded area representing the standard deviation. (e-f) Results from objects with random motions. (e) Depth color-coded MIP of a reconstructed frame showcasing fluorescent beads undergoing randomly movements in a liquid

solution (see video of the moving beads in Visualization 1). (f) Zoom-in 3D volume renderings detail the random 3D trajectories of moving beads, as denoted by the dotted lines.

## 2.3 Imaging of dynamic blinking objects

In addition to capturing rapid motions, another important category of complex and dynamic biological processes entail rapidly blinking signals, such as those arising from neuronal activities. To evaluate EventLFM's capability of tracking these types of dynamic signals, we employ a high-power LED driver (DC2200, Thorlabs) to generate adjustable pulsed illumination. In this proof-of-concept study, a 3D phantom embedded with fluorescent beads is illuminated using a variable pulse sequence, configured with a 1 ms pulse width and a variable pulse repetition rate ranging from 2 ms to 50 ms.

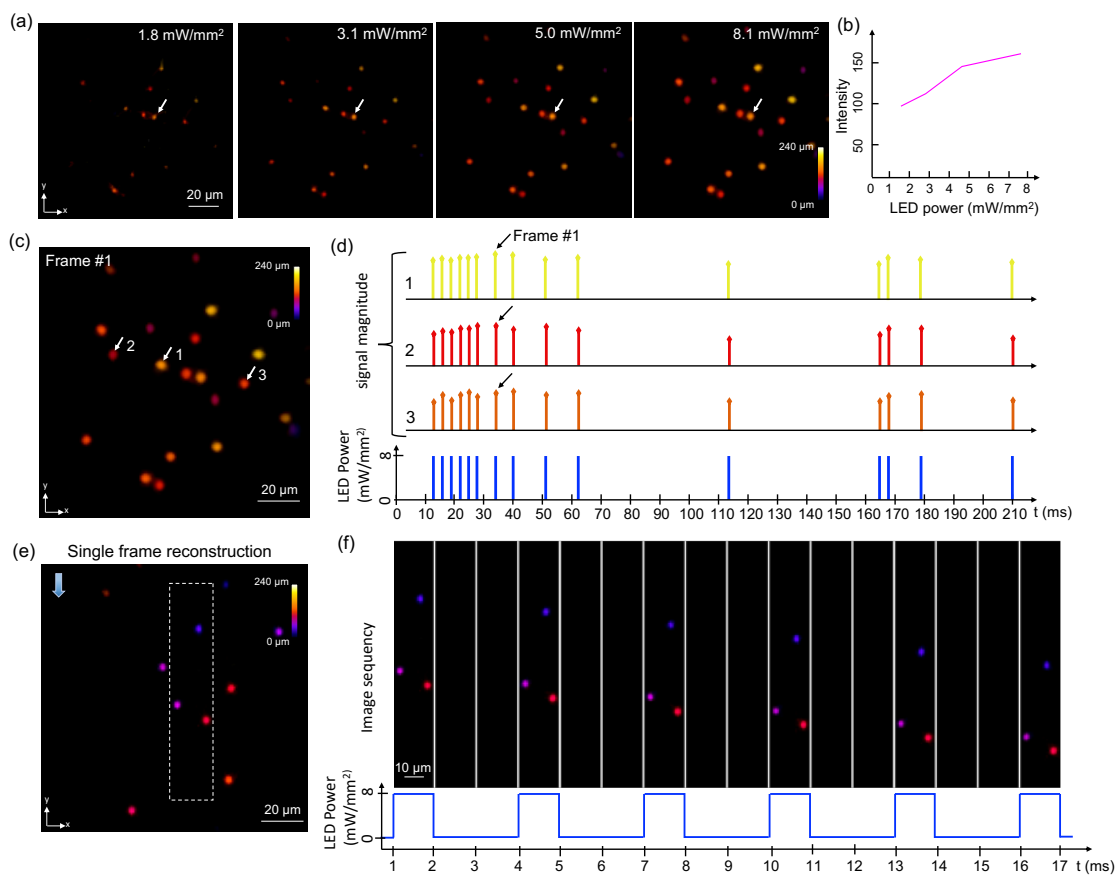

**Figure 4: EventLFM imaging of dynamic blinking objects.** (a) Depth color-coded MIPs of single-frame reconstructions capturing a blinking object under variable LED illumination powers. The LED pulse width is set to 1 ms with an inter-pulse interval of 2 ms. (b) Quantitative analysis of the reconstructed bead intensity as a function of LED power. Intensity measurements are averaged over a region within a single bead, as indicated by an arrow in (a). (c) Depth color-coded MIP of a single frame extracted from the reconstructed 3D volume, showcasing blinking fluorescent beads. (d) Temporal trace analysis is performed by calculating the mean intensities of three distinct beads, labeled as 1, 2, and 3, from the reconstructed volume. The LED pulse widths are uniformly set at 1 ms, while the inter-pulse intervals vary randomly between 2 ms and 50 ms. (e) Depth color-coded MIP of a single frame from the reconstructed moving phantom embedded with blinking fluorescent beads. (f) A temporal sequence of depth color-coded MIPs from the white dashed rectangular region

in (e) demonstrates synchronization with the LED pulse sequence shown below. The reconstructed bead positions are in agreement with the input linear motion.

It should be noted that the event-based signal features from blinking objects differ from those of fast-moving objects. Thus, an additional system characterization tailored to blinking objects is carried out. In this experiment, we maintain a pulse width and accumulation time of 1 ms width, while the optical power when the LED is on is systematically altered between 1.8 mW/mm$^2$ and 8.1 mW/mm$^2$. As shown in Fig. 4(a), the reconstructed signal increases with the applied optical power. We further quantify this relationship by isolating a single bead and calculating its mean reconstructed intensity at various illumination powers; the resultant graph presented in Fig. 4(b) reveals an approximately linear relationship.

Next, we demonstrate EventLFM's ability to image 3D objects blink at disparate intervals. For post-processing, a 1 ms accumulation time (equivalent to a 1 kHz frame rate) is set, synchronized to the onset of the first pulse. By employing the light field refocusing algorithm, we successfully reconstruct the blinking beads as displayed in Fig. 4(c). To further validate the system's accuracy, three distinct beads (as marked in the MIP in Fig. 4(c)) are selected and their mean intensity signals calculated, as shown in Fig. 4(d). The temporal traces confirm that the reconstructed signals, despite slight fluctuations in intensities, are in agreement with the pre-configured LED pulse sequences. This result validates EventLFM's capability in capturing high-frequency blinking signals in a 3D spatial context.

To provide a comprehensive assessment, we introduce concurrent linear motion to the blinking objects by synchronizing pulsed illumination with translational movement of the 3D phantom via a motorized stage. A phantom embedded with fluorescent beads is used similar to earlier experiments. Parameters are also set similar to earlier experiments, with a pulse width of 1 ms and a 2 ms delay, while the object velocity is fixed at 2.5 mm/s. Fig. 4(e) shows a depth color-coded MIP from a single reconstructed volume frame. To elucidate the dynamic objects further, Fig. 4(f) illustrates 16 consecutive frames within the white dashed rectangular region indicated in Fig. 4(e). These frames clearly show the expected linear motion, and the blinking events are reconstructed at the expected timestamps. Each reconstructed bead is translated linearly along the *y*-axis, as expected. Each signal-bearing frame is followed by two empty frames, which conform to the set LED pulse sequences shown in the bottom panel of Fig. 4(f). These results confirm EventLFM's robust and reliable performance in capturing complex 3D dynamics.

## 2.4 Imaging of neuronal signals in scattering mouse brain tissue

To demonstrate EventLFM's potential for neural imaging, we image a 75 μm thick section of GFP-labeled mouse brain tissue. Initially, the sample is uniformly illuminated using a pulsed LED source with a 1 ms pulse width. To validate the spatial reconstruction accuracy of EventLFM, we capture the fluorescence signals with traditional Fourier LFM and conventional fluorescent microscopy under constant illumination. Fig. 5(a) shows MIPs from a single reconstructed frame of each method. By visual inspection, the reconstruction from EventLFM is consistent with Fourier LFM, effectively capturing all neurons within the FOV and the intensity variations among them. However, a notable difference arises in the signal-to-background ratio (SBR). Fourier LFM suffers from a low SBR due to tissue scattering, which results in neuronal signals being buried in strong background fluorescence. In contrast, EventLFM demonstrates a significantly improved SBR, yielding a reconstruction with markedly improved image contrast and suppressed background fluorescence. This improvement is attributed to the event-based measurement mechanism intrinsic to EventLFM, wherein a readout is generated only when intensity changes exceed a certain threshold. Consequently, temporally slowly varying background fluorescence signals, which do not often meet this criterion, are either removed or substantially reduced in the raw data.

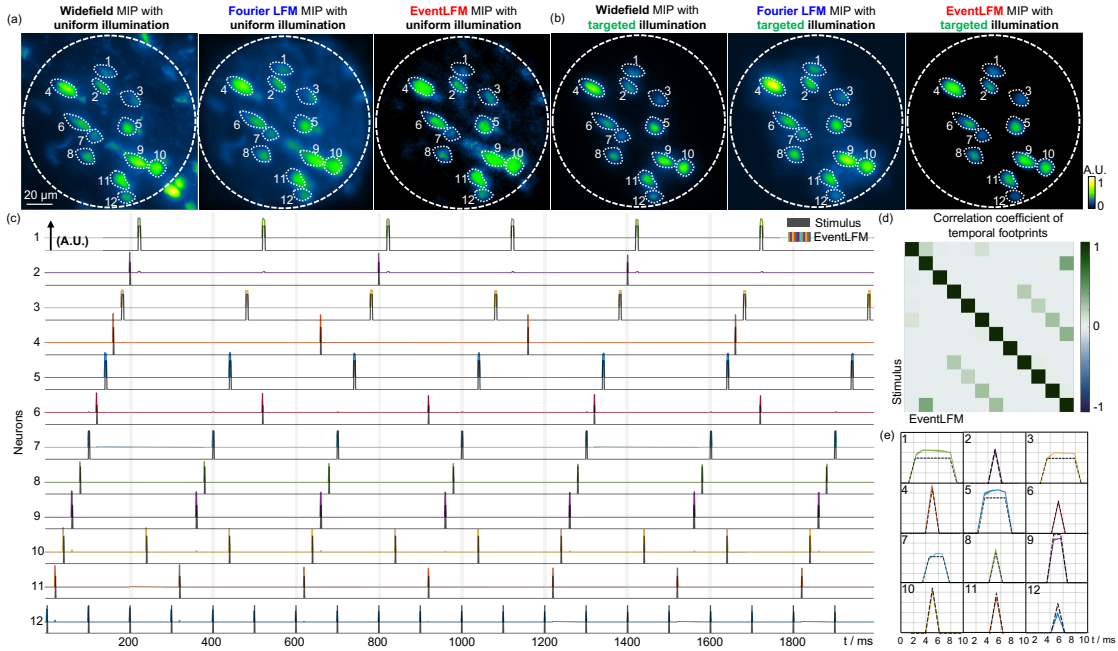

**Figure 5: EventLFM imaging of neuronal activities in mouse brain tissue.** (a, b) Spatiotemporal MIP comparison between axial scanned widefield microscopy, Fourier LFM, and EventLFM under uniform and targeted illumination. (c) Simulated neuronal activities within mouse brain tissue across various frequencies and time constants using a sequence of pre-modulated DMD patterns. Black temporal traces illustrate the stimulus from DMD-modulated illumination targeting twelve distinct neurons labeled in (a, b), while colored temporal traces represent the recorded neuronal signals, derived from averaging the intensities for each neuron. (d) Normalized cross-correlations of the neuronal traces extracted from EventLFM recording and their sorted ground-truth stimuli. (e) Averaged intensity (line) and standard error (shaded area) across the spike train for each neuron, the averaged pulse width accurately matching the intensity-normalized stimulus (dotted line).

Given that neuronal activity is often characterized by temporal intensity variations with minimal spatial movements, we implemented a targeted illumination strategy with pre-designed DMD patterns to further reduce the background noise [37]. These patterns modulate the excitation light to selectively illuminate the neurons, thereby mitigating the undesired background fluorescence excited from other sample regions. Moreover, to simulate realistic complex neuronal activities [15, 25, 38], we modulate a series of DMD patterns with various widths ranging from 1 ms to 6 ms and intervals from 100 ms to 600 ms, generating unique spike trains for each neuron. Initially, to illustrate targeted illumination's efficiency in background suppression, Fig. 5(b) provides the spatiotemporal MIPs of the time-series measurements from the widefield microscopy, Fourier LFM, and EventLFM with targeted illumination. By employing pre-designed DMD patterns (labeled by white dashed lines), only neurons are selectively illuminated within the brain slice, therefore providing a much cleaner background compared with the corresponding MIPs from the uniform illumination shown in Fig. 5 (a). When comparing the MIPs obtained from EventLFM and Fourier LFM, EventLFM again demonstrates improved signal contrast and background suppression capability by leveraging its event-driven measurements.

Additionally, to highlight EventLFM's capability in capturing rapid neuronal activities, Fig. 5(c) presents temporal traces from 12 distinct neurons, which are closely aligned with the pre-modulated DMD stimulus. Moreover, Fig. 5(d) presents a quantitative evaluation through the normalized correlation coefficient between neuronal signals extracted from EventLFM and the corresponding sorted ground-truth stimulus over a time duration spanning 2.4 s. The diagonal elements reaching a value of '1' underscore the strong correlation between temporal traces extracted from EventLFM and their respective ground truths. Non-diagonal elements reveal the presence of harmonic frequencies in the stimulus and potential signal crosstalk among closely situated neurons. These alignments and strong correlations validate EventLFM's capability to accurately reconstruct neuronal blinking dynamics within scattering tissues. Finally, utilizing predefined pulse widths and intervals for the spike trains of 12 neurons, we determine the temporal locations and extract all spikes within a 2-second duration from reconstructed traces. Fig. 5(e) presents the averaged intensity and standard error of the extracted single spikes for all 12 neurons. For validation, these spikes are compared with the ground truth stimulus, where the amplitude is defined by the averaged fluorescence intensity for each neuron. The pulse width derived from EventLFM reconstruction matches the stimulus precisely, demonstrating EventLFM's capability to robustly and accurately capture the unique footprint for complex neuronal dynamics. Additional details including DMD pattern generation, denoising of the event stream from the brain slice measurements, 3D reconstruction results of the brain slice, and additional experimental results with pulsed illumination can be found in Section 8 and Section 9 of Supplement 1.

### 2.5    Imaging of neuron-labeled freely moving *C. elegans*

To further showcase EventLFM's ability to capture complex biological dynamics, we employ it to track GFP-labeled neurons in a sample containing multiple *C. elegans*. For the experiment, the *C. elegans* are positioned on a gel substrate and subsequently submerged in a droplet of S-Basal solution, thereby creating a 3D environment for their free movement. First, we identify four distinct GFP-expressing neurons using conventional fluorescent microscopy -- two located in the tail region and another two in the mid-body section, as visualized in Fig. 6(a). Despite the relative sparsity of neurons, multiple *C. elegans* specimens are placed within the FOV. To accumulate enough event data for weaker neuronal signals, we set the accumulation time at 2 ms, yielding an effective frame rate of 500 Hz, which is sufficient for real-time 3D tracking of the organism. Using our EventLFM reconstruction algorithm, we generate depth color-coded MIP of the reconstructed volume frame at time 0 ms in Fig. 6(b), which clearly shows the spatial distribution of neurons across different depths for four distinct *C. elegans*. To further extract the neuronal dynamics, we focus on a specific region marked by a white dashed rectangle in Fig. 6(b). Two temporally separated 3D reconstructions from this region are presented at timestamps 42 ms and 92 ms in Fig. 6(c), complete with tracked trajectories marked in dashed lines. To further examine the neuronal movements, we present a time-series montage of the aforementioned area in Fig. 6(d) (additional results and comparisons with standard Fourier LFM are shown in Section 7 of Supplement 1). Notably, the neurons displayed in blue exhibit rapid and ascending motion across multiple axial planes over the time course. These result showcase EventLFM's capability to accurately capture biological dynamics in a 3D space at ultra-high frame rates.

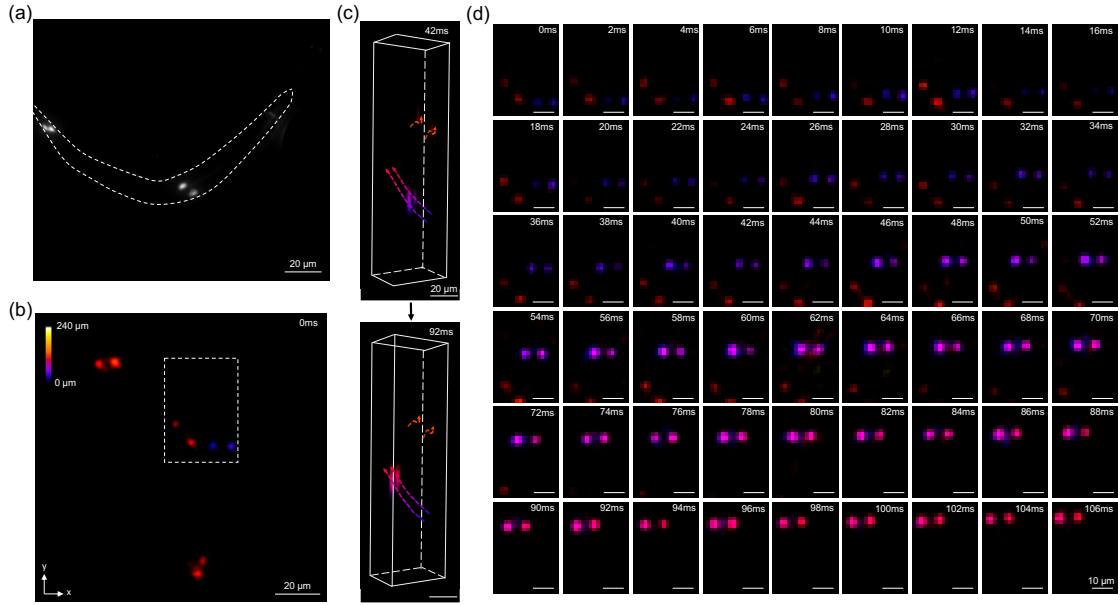

**Figure 6: EventLFM imaging of freely moving *C. elegans*.** (a) Representative image of *C. elegans* specimen captured via conventional widefield fluorescence microscopy, with the organism's contour delineated by the dashed line. (b) MIP of a depth color-coded reconstructed volume. (c) Time-resolved 3D reconstructions extracted from the white dashed rectangle region in (b) at timestamps 42 ms and 92 ms. The tracked trajectories over a 106 ms sequence are shown in the dashed lines. (d) Sequential MIPs extracted along the white dashed rectangle in (b), spanning a duration of 106 ms, reveal the 3D motions of neurons within the *C. elegans* organism.

## 2.6   Reconstruction results with deep learning

While the light-field refocusing algorithm is efficient, straightforward, and robust to various samples, it is susceptible to ghost artifacts and axial elongations in the 3D reconstructions. To address these issues, we implemented a convolutional neural network (CNN), modified from our previously developed CM²Net [9], tailored for high-resolution volumetric reconstruction in imaging systems with multi-view geometry. This network is trained with an experimentally collected dataset containing 3D moving particle phantoms. Event streams from these phantoms are integrated to generate the time surface frame at 1 ms intervals, which are then cropped to 5 x 5 view stacks, and generated refocused volumes

(RFVs) as inputs for network reconstruction. Corresponding ground-truth volumes are established through axial scans from conventional wide-field microscopy (20X, 0.75 NA).

Initially, the trained network is tested on a fast-moving 3D phantom containing 2 μm fluorescent particles. Fig. 7(a) illustrates a depth color-coded comparison of CNN reconstruction, RFV, and ground-truth volume, revealing that both CNN and RFV accurately recover the 3D information, with the CNN offering enhanced 3D resolution and suppressing the refocusing artifacts compared to RFV. The improvement in 3D resolution is quantified by comparing the *x* and *z* profiles of a single bead, shown in Fig. 7(b), where the CNN's profile aligns closely with the widefield system, demonstrating a ~2x enhancement over the RFV. This resolution enhancement is attributed to the network's ability to leverage subpixel parallax shifts across various MLA views, allowing for additional spatial information not accessible through conventional single-lens views. To further demonstrate the network's ability to accurately capture continuous motion, Fig. 7(c) presents a time color-coded 3D reconstruction and ground truth volume spanning 100 ms. The motion is further visualized by extracting three representative depths at the top, center, and bottom planes within the 3D volumes. The trajectories of particle motions (colored straight line), sampled at 1 kilohertz rate, verify the network's ability to robustly track the fast-moving objects with high 3D resolution.

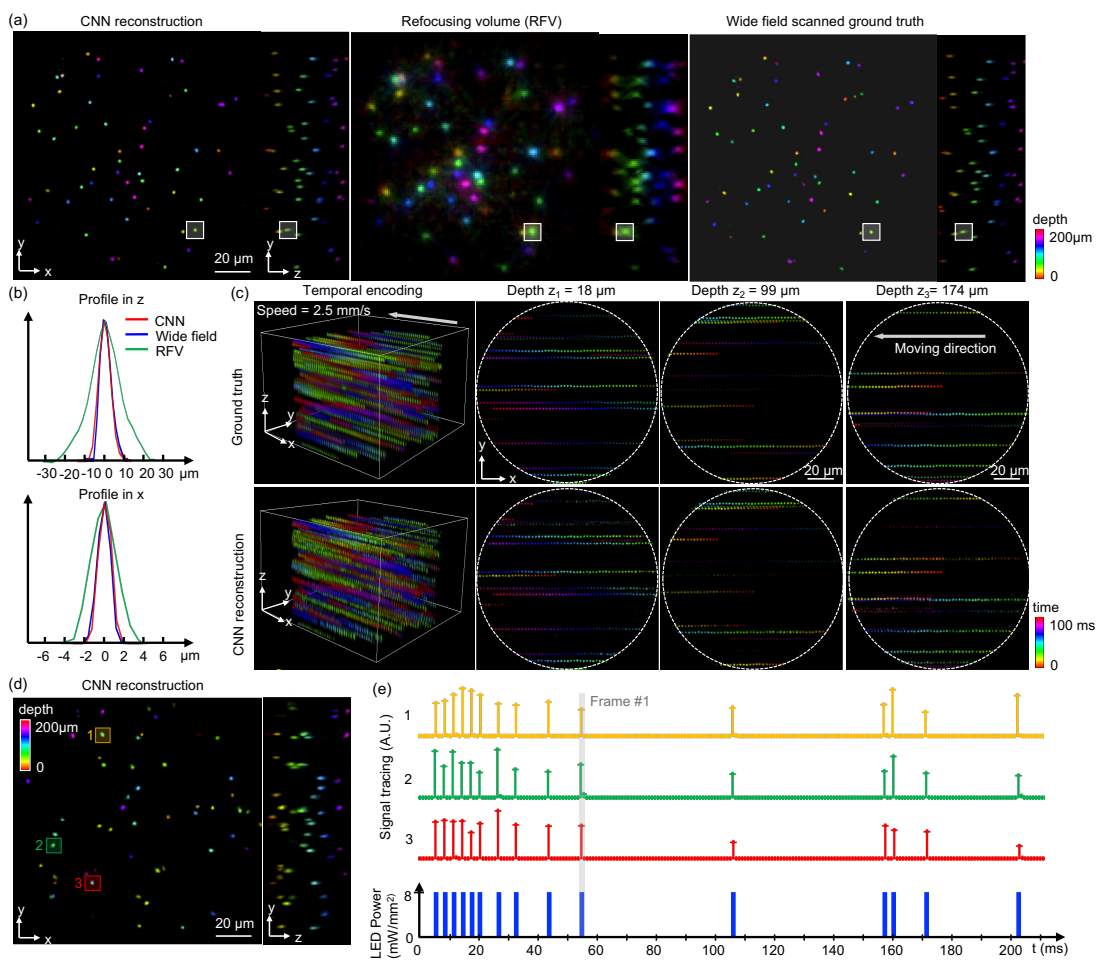

**Figure 7: Deep learning reconstruction results for EventLFM.** (a) Comparison between CNN reconstruction, RFV, and the ground-truth obtained via wide field microscopy with axial scanning (20x 0.75 NA). (b) Axial and lateral profile of a single bead, as indicated by the white dashed line in (a). The CNN reconstruction (red line) improves both axial and lateral resolution compared to RFV (green line) and matches the particle size as validated by the wide field system. (c) Visualization of scanned wide field ground-truth volume and 3D reconstructions across 100 sequential frames with time-coded by color. Three representative depths at the top, center, and bottom within the 3D volume are extracted for close validation. The trajectories of rapidly moving particles, sampled at 1kHz rate, are delineated by straight lines. (d) Depth color-coded 3D CNN reconstruction of a frame marked as Frame #1. (e) Temporal trace analysis for three particles, labeled 1, 2, and 3, within the 3D reconstruction. The LED pulse widths are uniformly set at 1 ms with the inter-pulse intervals randomly varying between 2 ms and 50 ms.

Finally, we conduct an experiment to evaluate the network's performance on dynamic blinking objects. Fig. 7(d) presents a depth-color-coded CNN reconstruction of a single frame, demonstrating the high 3D resolution capability in such conditions. Despite the differing event characteristics between blinking and moving objects, employing the RFV as an initial estimation allows the network to effectively generalize across both types of samples. Additionally, Fig. 7(e) illustrates the temporal traces of three particles within the reconstruction, which precisely align with the LED illumination pulse sequence. This demonstrates the network's ability to accurately capture high-frequency blinking signals in a 3D context at enhanced resolutions using the CNN.

These preliminary reconstruction results on both fast-moving and dynamic-blinking phantoms demonstrate that EventLFM augmented by deep learning enables ultrafast 3D imaging with high 3D resolution and minimal reconstruction artifacts.

## 3 Discussion

We present the first, to the best of our knowledge, ultrafast Fourier LFM system, EventLFM, that leverages an event camera and a tailored reconstruction algorithm to facilitate volumetric imaging at kHz speeds. By comparing the PSF, MTF and 3D reconstructions, we have established that EventLFM achieves a lateral resolution comparable to that of traditional Fourier LFM system. Notably, EventLFM provides marginally superior axial resolution and substantially improved temporal resolution. Our experimental results further underscore EventLFM's versatility and capability. We demonstrate its effectiveness in reconstructing complex dynamics of rapidly moving 3D objects at 1 kHz temporal resolution. Moreover, through controlled illumination experiments, we showcase imaging of high-frequency 3D blinking objects with pulse widths as short as 1 ms. Additionally, we demonstrate EventLFM's ability to capture rapid dynamic signals within scattering tissues by imaging realistic neuronal activities in a mouse brain section simulated by a series of DMD patterns to induce unique spatiotemporal footprints at kHz rates. Moreover, we present imaging and tracking of GFP-expressing neurons in freely moving *C. elegans* within a 3D space, achieving a frame rate of 500 Hz. Lastly, we show the integration of a deep learning reconstruction network with EventLFM to improve the imaging quality and enhance the 3D resolution.

EventLFM, leveraging its unique capability to capture brightness change, augmented by targeted illumination, significantly mitigates the low SBR challenges typically encountered in scattering environments [39] - a major limitation of traditional widefield microscopy techniques - as demonstrated by our experiment on mouse brain tissues. However, the effectiveness of targeted illumination is currently constrained to shallow depth because it relies on a high NA objective to image the DMD patterns onto the sample plane. Future advancements to EventLFM can involve 3D targeted illumination techniques [40], potentially extending the SBR enhancement across the entire volume with improved 3D optical sectioning capability. In addition, our work opens tremendous opportunities for future research in event-driven imaging within scattering media [41] and the development of advanced computational algorithms that more effectively leverage event-driven measurement for extracting dynamic signals from deep within scattering tissues.

The high sensitivity of event cameras often results in increased noise levels, presenting challenges for their applications in microscopy settings. Our light field refocusing algorithm for EventLFM leverages the uncorrelated noise characteristic across different views under MLA to effectively reduce noise while amplifying the signal of in-focus signals through the shift-and-sum operation, resulting in an SNR improvement by 5x (An additional noise and SNR analysis is provided in Section 4 of Supplement 1). For future developments, there is significant potential for employing advanced denoising algorithms to further suppress stochastic event sensor noise [42, 43].

Our pilot study of deep learning for EventLFM reconstructions has yielded promising results, notably enhancing 3D spatial resolution and mitigating the ghost artifacts inherent to the refocusing algorithm. Nonetheless, the current training dataset collection process is both time-consuming and suffers from limited data diversity due to the difficulty of acquiring a large-scale ground truth high-speed (e.g. kHz), high-resolution volumes on diverse biological samples. A promising future direction is to develop a physics-based simulator that can efficiently generate a broad range of training data. This will significantly enhance the network's ability to generalize to more complex biological contexts. Furthermore, the simulator will pave the way for investigating more advanced deep learning techniques, such as employing a demultiplexing network for expanding the imaging FOV [44], thus compensating for the event camera's limited pixel array constraints. Another possible direction involves developing an adaptive encoder to enhance the extraction of physically meaningful information from the sparse event data [42, 43] to fully exploit the sparsity of event data [45] to reconstruct more complex 3D processes over large volumes.

In conclusion, given its simplicity, ultrafast 3D imaging capability, and robustness in scattering environments, EventLFM has the potential to be a valuable tool in various biomedical applications for visualizing complex, dynamic 3D biological phenomena.

## 4  Materials and methods

### 4.1  Experimental setup

EventLFM augments a conventional Fourier LFM setup with an event camera (EVK4, Prophesee, IMX636 sensor, 1280 X 720 pixels, 4.86 μm pixel size), as shown in Fig. 1(a). A blue LED (SOLIS-470C, Thorlabs) serves as the excitation illumination source for the fluorescent samples. This excitation light is focused onto the back pupil plane of the objective lens (Plan Apo, 20X, 0.75 NA, Nikon) to ensure uniform illumination across the target volume. In addition, we add another branch in the illumination path that includes a DMD (DLI4130 0.7 XGA VIS High-Speed Kit, Digital Light Innovations) for manipulating the spatiotemporal distribution of the excitation light within the FOV. In the detection path, fluorescence emissions from the sample are collected by the objective lens and subsequently relayed to an intermediate image plane by a tube lens (TL, $f$ = 200 mm, ITL200, Thorlabs). This intermediate image is then transformed by a Fourier lens (FL, $f$ = 80 mm, AC508-080-A, Thorlabs). An MLA ($f$ = 16.8 mm, S600-f28, RPC Photonics) is placed at the back focal plane of the FL to achieve uniform angular sampling, thereby generating a 5 X 5 subimage array. An ancillary 4f relay system ($M$ = 1.25, not shown) is placed after the MLA to ensure optimal distribution of these subimages across the event sensor. In addition, a 50/50 beamsplitter is integrated within the 4f system to enable simultaneous capture of the dynamic 3D volumes by both the event camera and an sCMOS camera (CS2100M-USB, Thorlabs), thereby providing a direct comparative benchmark between EventLFM and traditional Fourier LFM modalities. Further details can be found in Section 1 of Supplement 1.

## 4.2 Reconstruction algorithm

The event camera records the polarity of changes in pixel intensity as an event stream with a temporal granularity as fine as 1 μs. Specifically, an event is generated when a dynamic or luminous variation within the FOV surpasses a pre-defined threshold. For each event, the sensor outputs the spatial coordinate x and y, the precise timestamp t, and the polarity p (either positive or negative depending on the direction of the intensity change). This asynchronous event stream is then sorted based on the polarity and integrated within a user-defined accumulation time to construct temporally continuous frames, as shown in Fig. 1(b). Our sensor has a pixel latency of 100 μs - 220 μs, setting the upper limit for accumulation time. To ensure enough events for robust frame reconstruction, we set this time window between 1 to 2 milliseconds based on the specific sample under investigation. Importantly, the chosen accumulation time directly determines the system's frame rate while also affecting the reconstruction resolution, details of which are elaborated in Section 6 of Supplement 1.

Rather than simply summing the events within the accumulation period, we apply a built-in time-surface algorithm for the post-processing of the raw event data. This algorithm employs an exponential time-decay function to compute a time surface, encapsulating both the spatial and temporal correlations among adjacent pixels. Pixel values in this time surface, ranging from 0 to 255, are indicative of historical temporal activity, as illustrated in Fig. 1(c). This approach offers a spatiotemporal representation for each event while mitigating motion blur artifacts (see Section 3 of Supplement 1 for details). In addition, we apply median filtering (*medfilt2* in MATLAB) to suppress the sensor noise. Subsequently, these time-surface frames are processed by a standard light field refocusing algorithm [36] to yield a 3D volumetric reconstruction, as shown in Fig.1(d). To enhance the quality of the reconstruction, we either apply a predefined threshold or a deep neural network to remove ghosting artifacts introduced by the refocusing algorithm. For visualization, we opt for either depth- or time-encoding color schemes when appropriate, as in Fig. 1(d) and 1(e). A detailed reconstruction diagram can be found in Section 2 of Supplement 1.

The deep neural network is a modified CM²Net [9], which is developed to enhance the spatial resolution and improve the quality of the reconstruction. For the network training, we collect event streams of 200 μm thick phantoms embedded with 2 μm fluorescent particles moving at 2.5 mm/s. Corresponding ground-truth volumes are obtained by axial scanning using conventional wide-field microscopy (20X, 0.75 NA) at 3 μm step size, resulting in a total of 4700 training pairs. The training uses a loss function composed of the Normalized Pearson Correlation Coefficient (NPCC) and Mean Absolute Error (MAE):

$$L_{total} = L_{MAE} - L_{NPCC}, \tag{1}$$

$$L_{MAE} = \frac{1}{n}\sum_{i=1}^{n}|y_i - x_i|, \tag{2}$$

$$L_{NPCC} = \frac{\sum_{i=1}^{n}(y_i-\bar{y})(x_i-\bar{x})}{\sqrt{\sum_{i=1}^{n}(y_i-\bar{y})^2}\sqrt{\sum_{i=1}^{n}(x_i-\bar{x})^2}} \tag{3}$$

where $y_i$ and $x_i$ denote the true and predicted values, and $\bar{y}$ and $\bar{x}$ represent their respective averaged values, with i indexing the pixels and n representing the total number of pixels. This dual loss is designed to simultaneously enhance spatial alignment and reduce intensity discrepancies. The network is implemented with PyTorch and runs on an Nvidia GPU RTX 4090, with a batch size of 8. The entire training process takes 24 hours. The detailed network structure and implementations are provided in Section 10 of Supplement 1.

### 4.3 Preparation of brain slice

The preparation of rodent brain slices exhibiting GFP-tagged neurons within the bed nucleus of the stria terminalis (BNST) involved a series of detailed procedures. Initially, male C57Bl/6 mice were anesthetized with a steady inhalation of 2% isoflurane and received preemptive pain relief through buprenorphine and ketoprofen (0.5 and 5 mg/kg respectively). Then their heads were fixed within a digital stereotaxic apparatus (David Kopf Instruments, Tujunga, CA, USA). Following a midline incision on the skull and a small craniotomy bilaterally over each injection site, a viral vector carrying GFP (100-200 nL, pAAV-CAG-GFP, Addgene #37825-AAVrg) was injected into the BNST using a glass micropipette coupled with a Nanoject II injector (Drummond), targeting specific brain coordinates 1.0 mm lateral from the midline, 0.4 mm anterior to bregma, and a depth of 4.3 mm from dura. The retrograde virus labeled both local neurons at the injection site in BNST as well as upstream brain areas. Post-surgery, the mice were allowed a recovery period of two to three weeks for the virus to manifest before being deeply sedated for the final procedure. Their brains were then fixed in a paraformaldehyde solution, cryoprotected, and sectioned coronally at 75 μm thickness using a cryostat. These sections were finally mounted on Superfrost Plus slides (Fisher) and Vectashield mounting medium (Vector Labs).

This study was performed in strict accordance with the recommendations in the Guide for the Care and Use of Laboratory Animals of the National Institutes of Health. All animals were handled according to approved Institutional Animal Care and Use Committee (IACUC) protocols (#201800540) of Boston University.

### 4.4 Preparation of fluorescent beads phantom

To prepare the fluorescent particle phantom, we accurately pipette 5 µL of fluorescent beads (2 µm diameter, 1% concentration, Fluoro-Max Dyed Green Aqueous Fluorescent Particles) into 2 ml of clear resin (Formlabs, catalog no. RS-F2-GPCL-04), contained within a suitable tube. The mixture is then homogenized by ultrasonic probe sonicator (Fisherbrand™ Model 50 Sonic Dismembrator) to ensure a uniform distribution of beads within the resin. Subsequently, the solution is carefully dispensed into a rectangular mold, designed with a 200 µm depth, until the mold is filled. A glass slide is then placed over the mold to cover it. The resin-bead solution is solidified by exposing it to UV light. Finally, the glass slide is removed to retrieve a uniform phantom with a precise thickness of 200 µm.

### 4.5 Preparation of *C. elegans*

The transgenic *Caenorhabditis elegans* (*C. elegans*) strain ZB4510 [mec-4p::GFP], genetically modified to express green fluorescent protein (GFP) [46], are cultivated on nematode growth medium (NGM) agarose plates. The plates are coated with Escherichia coli to provide a consistent food source. We adhere to a strict subculturing routine, transferring the *C. elegans* to fresh media every three days. Before imaging, a 20 µL layer of 2% agarose solution (Sigma-Aldrich) is prepared on a 15 mm by 15 mm section of a glass microscope slide to form a pad. A small population of *C. elegans* is then transferred onto this agarose pad, and a drop of S-Basal is applied to facilitate free movement of the *C. elegans* within the solution. The prepared slide is immediately placed on the stage of the microscope for imaging.

## Data availability

Data underlying the results presented in this paper may be obtained from the authors upon reasonable request.

## Acknowledgements

The authors acknowledge funding from National Institutes of Health (R01NS126596) and a grant from 5022 - Chan Zuckerberg Initiative DAF, an advised fund of Silicon Valley Community Foundation. The authors acknowledge Danchen Jia and Dr. Ji-Xin Cheng for generously lending us the microlens array, Dr. Kevin J. Monk, Brett T. DiBenedictis, and Ian G. Davison for providing the brain slices, as well as Boston University Shared Computing Cluster for proving the computational resources.

## Conflict of interests

The authors declare no competing interests.

# Supplementary Information

See Supplement 1 for supporting content.

# References

[1] Mertz, J. Strategies for volumetric imaging with a fluorescence microscope. *Optica* **6**, 1261–1268 (2019).

[2] Minsky, M. Memoir on inventing the confocal scanning microscope. *Scanning* **10**, 128–138 (1988).

[3] Helmchen, F. & Denk, W. Deep tissue two-photon microscopy. *Nature Methods* **2**, 932–940 (2005).

[4] Voie, A. H., Burns, D. & Spelman, F. Orthogonal-plane fluorescence optical sectioning: Three-dimensional imaging of macroscopic biological specimens. *J. microscopy* **170**, 229–236 (1993).

[5] Levoy, M., Ng, R., Adams, A., Footer, M. & Horowitz, M. Light field microscopy. in Acm Siggraph 2006 Papers, pp. 924–934 (2006).

[6] Guo, C., Liu, W., Hua, X., Li H., & Jia, S. Fourier light-field microscopy. *Optics Express* **27**, 25573 (2019).

[7] Llavador, A., Sola-Pikabea, J., Saavedra, G., Javidi, B., & Martínez-Corral, M. Resolution improvements in integral microscopy with Fourier plane recording. *Optics Express* **24**, 20792–20798 (2016).

[8] Xue, Y., Davison, I. G., Boas, D. A. & Tian, L. Single-shot 3d wide-field fluorescence imaging with a computational miniature mesoscope. *Science Advances* **6**, eabb7508 (2020).

[9] Xue, Y., Yang, Q., Hu, G., Guo, K., & Tian, L. Deep-learning-augmented computational miniature mesoscope. *Optica* **9**, 1009–1021 (2022).

[10] Liu, F. L., Kuo, G., Antipa, N., Yanny, K. & Waller, L., Fourier diffuserscope: single-shot 3d fourier light field microscopy with a diffuser. *Optics Express* **28**, 28969–28986 (2020).

[11] Adams, J. K., Boominathan, V., Avants, B. W., Vercosa, D. G., Ye, F., Baraniuk, R. G., Robinson, J. T. & Veeraraghavan, A. Single-frame 3d fluorescence microscopy with ultraminiature lensless flatscope. *Science Advances* **3**, e1701548 (2017).

[12] Nelson, S. & Menon, A. Bijective-constrained cycle-consistent deep learning for optics-free imaging and classification. *Optica* **9**, 26–31 (2022).

[13] Nehme, E., Freedman, D., Gordon, R., Ferdman, B., Weiss, L. E., Alalouf, O., Naor, T., Orange, R., Michaeli, T. & Shechtman. Y. Deepstorm3d: dense 3d localization microscopy and psf design by deep learning. *Nature Methods* **17**, 734–740 (2020).

[14] Pavani, S. R. P., Thompson, M. A., Biteen, J. S., Lord, S. J., Liu, N., Twieg, R. J., Piestun, R. & Moerner, W. E. Three-dimensional, single-molecule fluorescence imaging beyond the diffraction limit by using a double-helix point spread function. Proceedings of the National Academy of Sciences **106**, 2995–2999 (2009).

[15] Abdelfattah, A. S., Zheng, J., Singh, A., Huang, Y.-C., Reep, D., Tsegaye, G., Tsang, A., Arthur, B. J., Rehorova, M. *et al.* Sensitivity optimization of a rhodopsin-based fluorescent voltage indicator. *Neuron* **111**, 1547–1563.e9 (2023).

[16] Bouchard, M. B., Chen, B. R., Burgess, S. A. & Hillman, E. M. Ultra-fast multispectral optical imaging of cortical oxygenation, blood flow, and intracellular calcium dynamics. *Optics Express* **17**, 15670–15678 (2009).

[17] Rome, L. C. & Lindstedt, S. L. The quest for speed: muscles built for high-frequency contractions. *Physiology* **13**, 261–268 (1998).

[18] Gao, L., Liang, J., Li, C. & Wang, L. V. Single-shot compressed ultrafast photography at one hundred billion frames per second. *Nature* **516**, 74–77 (2014).

[19] Liang, J. & Wang, L. V. Single-shot ultrafast optical imaging. *Optica* **5**, 1113–1127 (2018).

[20] Liu, X., Skripka, A., Lai, Y., Jiang, C., Liu, J., Vetrone, F. & Liang, J. Fast wide-field upconversion luminescence lifetime thermometry enabled by single-shot compressed ultrahigh-speed imaging. *Nature Communications* **12**, 6401 (2021).

[21] Ma, Y., Lee, Y., Best-Popescu, C. & Gao, L. High-speed compressed-sensing fluorescence lifetime imaging microscopy of live cells. *Proc. National Acad. Sci.* **118**, e2004176118 (2021).

[22] Feng X. & Gao, L. Ultrafast light field tomography for snapshot transient and non-line-of-sight imaging. *Nature Communications* **12**, 2179 (2021).

[23] Weber, T. D., Moya, M. V., Kılıç, K., Mertz, J. & Economo, M. N. High-speed multiplane confocal microscopy for voltage imaging in densely labeled neuronal populations. *Nature Neuroscience* **26**, 1642–1650 (2023).

[24] Wu, J., Liang, Y., Chen, S., Hsu, C. L., Chavarha, M., Evans, S. W., Shi, D., Lin, M. Z., Tsia, K. K. & Ji, N. Kilohertz two-photon fluorescence microscopy imaging of neural activity in vivo. *Nature Methods* **17**, 287–290 (2020).

[25] Platisa, J., Ye, X., Ahrens, A. M., Liu, C., Chen, I. A., Davison, I. G., Tian, L., Pieribone, V. A. & Chen, J. L. High-speed low-light in vivo two-photon voltage imaging of large neuronal populations. *Nature Methods* **20**, 1095–1103 (2023).

[26] Xiao, S., Giblin, J. T., Boas, D. A. & Mertz, J. High-throughput deep tissue two-photon microscopy at kilohertz frame rates. *Optica* **10**, 763–769 (2023).

[27] Gallego, G., Delbrück, T., Orchard, G., Bartolozzi, C., Taba, B., Censi, A., Leutenegger, S., Davison, A. J., Conradt, J., Daniilidis, K. & Scaramuzza, D. Event-based vision: A survey. *IEEE Transactions on Pattern Analysis and Machine Intelligence* **44**, 154–180 (2022).

[28] Lichtsteiner, P., Posch, C. & Delbruck, T. A 128× 128 120 dB 15 $\mu$s latency asynchronous temporal contrast vision sensor. *IEEE Journal of Solid-State Circuits* **43**, 566–576 (2008).

[29] Willert, C. E. Event-based imaging velocimetry using pulsed illumination. *Experiments in Fluids* **64**, 98 (2023).

[30] Chen, G., Cao, H., Conradt, J., Tang, H., Rohrbein, F. & Knoll, A. Event-based neuromorphic vision for autonomous driving: A paradigm shift for bio-inspired visual sensing and perception. *IEEE Signal Processing Magazine* **37**, 34–49 (2020).

[31] Amir, A., Taba, B., Berg, D., Melano, T., McKinstry, J., Nolfo, C. Di, Nayak, T., Andreopoulos, A., Garreau, G., Mendoza, M. *et al.* A low power, fully event-based gesture recognition system. *Proceedings of the IEEE conference on computer vision and pattern recognition*, 7243–7252 (2017).

[32] Cabriel, C., Monfort, T., Specht, C. G. & Izeddin, I. Event-based vision sensor for fast and dense single-molecule localization microscopy. *Nature Photonics* **17**, 1105–1113 (2023).

[33] Mangalwedhekar, R., Singh, N., Thakur, C. S., Seelamantula, C. S., Jose, M. & Nair, D. Achieving nanoscale precision using neuromorphic localization microscopy. *Nature Nanotechnology* **18**, 380–389 (2023).

[34] Lagorce, X., Orchard, G., Galluppi, F., Shi, B. E. & Benosman, R. B. HOTS: A hierarchy of event-based time-surfaces for pattern recognition. *IEEE Transactions on Pattern Analysis and Machine Intelligence* **39**, 1346–1359 (2017).

[35] Sironi, A., Brambilla, M., Bourdis, N., Lagorce, X. & Benosman, R. Hats: Histograms of averaged time surfaces for robust event-based object classification. *Proceedings of the IEEE Conference on Computer Vision and Pattern Recognition*, 1731-1740 (2018).

[36] Ng, R., Levoy, M., Brédif, M., Duval, G., Horowitz, M. & Hanrahan, P. Light field photography with a hand-held plenoptic camera. Ph.D. thesis, Stanford University (2005).

[37] Xiao, S., Lowet, E., Gritton, H. J., Fabris, P., Wang, Y., Sherman, J., Mount, R. A., Tseng, H. A., Man, H. Y., Straub, C., Piatkevich, K. D., Boyden, E. S., Mertz, J. & Han, X. Large-scale voltage imaging in behaving mice using targeted illumination. iScience **24**(11): 103263 (2021).

[38] Abdelfattah, A. S., Kawashima, T., Singh, A., Novak, O., Liu, H., Shuai, Y., Huang, Y. *et al.* Bright and photostable chemigenetic indicators for extended in vivo voltage imaging. Science **365**, no. 6454, 699-704 (2019).

[39] Mertz, J. Optical sectioning microscopy with planar or structured illumination. Nature Methods **8**, 811–819 (2011).

[40] Zhang, J., Pégard, N., Zhong, J., Adesnik, H. & Waller, L. 3D computer-generated holography by non-convex optimization. Optica **4**, 1306-1313 (2017)

[41] Zhang, N., Shea, T. & Nurmikko, A. Event-Driven Imaging in Turbid Media: A Confluence of Optoelectronics and Neuromorphic Computation. arXiv: 2309.06652 (2023).

[42] Zhang, Z., Suo & J., Dai, Q. Denoising of event-based sensors with deep neural networks. *Optoelectronic Imaging and Multimedia Technology VIII*. Vol. 11897. SPIE, (2021).

[43] Hagenaars, J., Paredes-Vallés, F. & De Croon, G. Self-supervised learning of event-based optical flow with spiking neural networks. *Advances in Neural Information Processing Systems* **34**, 7167–7179 (2021).

[44] Yang, Q., Guo, R., Hu, G., Xue, Y., Li, Y. & Tian, L. Wide-Field, High-Resolution Reconstruction in Computational Multi-Aperture Miniscope Using a Fourier Neural Network. arXiv:2403.06439 (2024).

[45] Eshraghian, J. K., Ward, M., Neftci, E. O., Wang, X., Lenz, G., Dwivedi, G., Bennamoun, M., Jeong, D. S. & Lu, W. D. Training spiking neural networks using lessons from deep learning. in Proceedings of the *IEEE* **111**, 1016-1054 (2023).

[46] Wang, G., Sun, L., Reina, C. P., Song, I., Gabel, C. V. & Driscoll, M. Ced-4 card domain residues can modulate non-apoptotic neuronal regeneration functions independently from apoptosis. *Scientific Reports* **9**, 13315 (2019).

# Supplementary information

## EventLFM: Event Camera integrated Fourier Light Field Microscopy for Ultrafast 3D imaging

Ruipeng Guo[1], Qianwan Yang[1], Andrew S. Chang[4], Guorong Hu[1], Joseph Greene[1], Christopher V. Gabel[3,4], Sixian You[5], and Lei Tian[1,2,3,*]

[1]Department of Electrical and Computer Engineering, Boston University, Boston, MA 02215, USA.
[2]Department of Biomedical Engineering, Boston University, Boston, MA 02215, USA.
[3]Neurophotonics Center, Boston University, Boston, MA 02215, USA.
[4]Department of Physiology and Biophysics, Boston University, Boston, MA 02215, USA.
[5]Research Laboratory of Electronics (RLE) in the Department of Electrical Science and Engineering, Massachusetts Institute of Technology, Cambridge, MA 02139, USA.

[*]Correspondence: leitian@bu.edu, Tel.: 1-617-353-1334

Authors' email:

Ruipeng Guo: rguo@bu.edu

Qianwan Yang: yaw@bu.edu

Andrew S. Chang: aschang@bu.edu

Guorong Hu: grhu@bu.edu

Joseph Greene: joeg18@bu.edu

Christopher V. Gabel: cvgabel@bu.edu

Sixian You: sixian@mit.edu

Lei Tian: leitian@bu.edu

## 1. Setup and system characterization

The EventLFM system is a novel integration of a traditional Fourier light field microscope (LFM) and an event camera, as shown in Fig. S1. Our system consists of three primary optical paths: the illumination path, the Fourier LFM imaging path and the reference path.

**Illumination path**: A blue LED (SOLIS-470C, Thorlabs) is used as the excitation source for imaging GFP fluorescence. To maximize light efficiency, two condensers (CL1 and CL2) are used to collect and collimate the highly divergent LED source. Next, two field lenses (L1 and L2) are used to focus the beam onto the back pupil plane of the objective lens (Plan Apo, 20×, Nikon) and pass through a filter set to ensure uniform illumination within the target volume.

**DMD setup**: A DMD (DLI4130 0.7 XGA VIS High-Speed Kit, Digital Light Innovations) is set up in illumination path to manipulate the distribution of excitation in FOV. Beam is illuminated onto the face of DMD with a degree of 24° through a TIR prism. Then the DMD face is imaged in the FOV with demagnification of ~40 by a 4F system and the objective. By designing the pattern on DMD, we can generate structured illumination as needed.

**Fourier LFM Detection Path**: Fluorescence emission from the sample is collected by the objective lens (Plan Apo, 20×, 0.75 NA, Nikon). A beamsplitter (BS028, Thorlabs) is placed after the filter set to split the fluorescence signals into the Fourier LFM and reference detection paths. For the Fourier LFM path, a tube lens (TL, f = 200 mm, ITL200, Thorlabs) projects the signal onto the focal plane of a Fourier lens (FL, AC508-080-A, Thorlabs), which performs a Fourier transform on the intermediate image. A microlens array (MLA, S600-f28, RPC photonics) is placed at the back focal plane of the FL to uniformly sample the angular information and generate a 5×5 array of elemental images. To efficiently utilize the sensor size, a 4f system is implemented after the MLA.

**Reference Path**: In parallel, the reference path employs a lens (LA1417-A, f = 150 mm, Thorlabs) and an sCMOS camera (CS2100M-USB, Thorlabs) to form a conventional widefield fluorescence microscope.

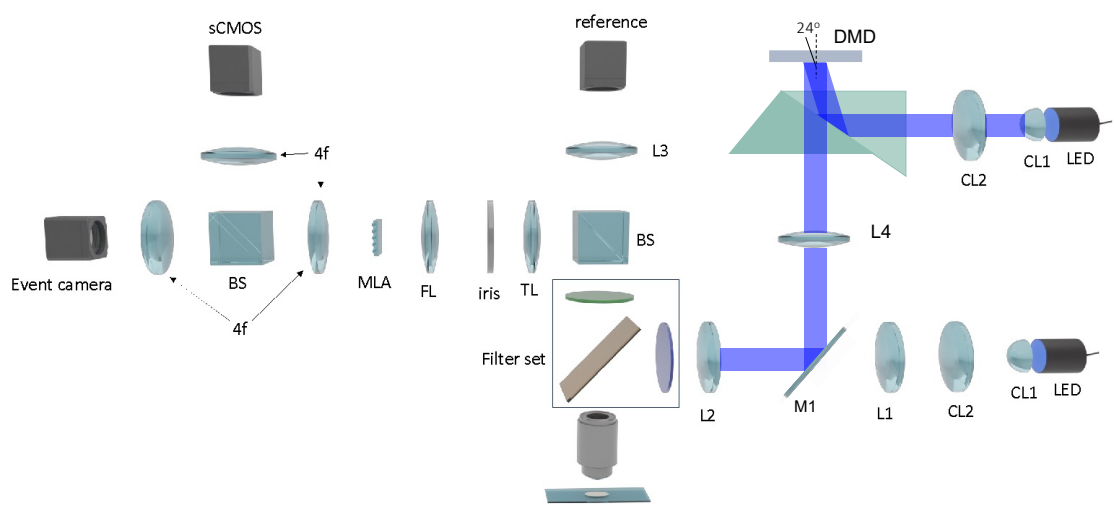

Figure S1: **Sketch of the EventLFM setup.**

**Integration of Event Camera**: Within the 4f system of the Fourier LFM path, an additional beamsplitter (BS013, Thorlabs) is introduced to distribute the light field onto both an event camera (EVK4, Prophesee) and an sCMOS camera (CS2100M-USB, Thorlabs). The 4f systems before the event camera and sCMOS camera share the same parameters. This configuration allows for simultaneous recording of dynamic fluorescence signals at identical magnifications, facilitating a direct and unbiased comparison between the EventLFM system and the traditional Fourier LFM.

We systematically assessed the performance characteristics of the traditional Fourier LFM system equipped with an sCMOS camera.

**Field of View (FOV) Calibration**: to determine the system's FOV, we employed a calibration strategy using a single reference bead. The bead was translated across the imaging plane using a motorized stage. By examining the maximum range of displacements without exceeding the boundaries of each elemental images, we quantified the FOV to be approximately 130 µm, as illustrated in Fig. S2(a).

**Depth of Field (DOF) Analysis**: Theoretically, our system's DOF is approximately 300 µm, based on the shift limit of images originating from the outermost microlenses. However, we observed severe aberrations when the bead is moved beyond the central 200 µm depth range. Figure S2(b) shows the images of a single bead at z-positions of -100 µm, 0 µm and 100 µm. Point spread function (PSF) aberrations are markedly evident at z-positions of ±100 µm from the focal plane. Consequently, we constrained the system's operational DOF to 200 µm in this study.

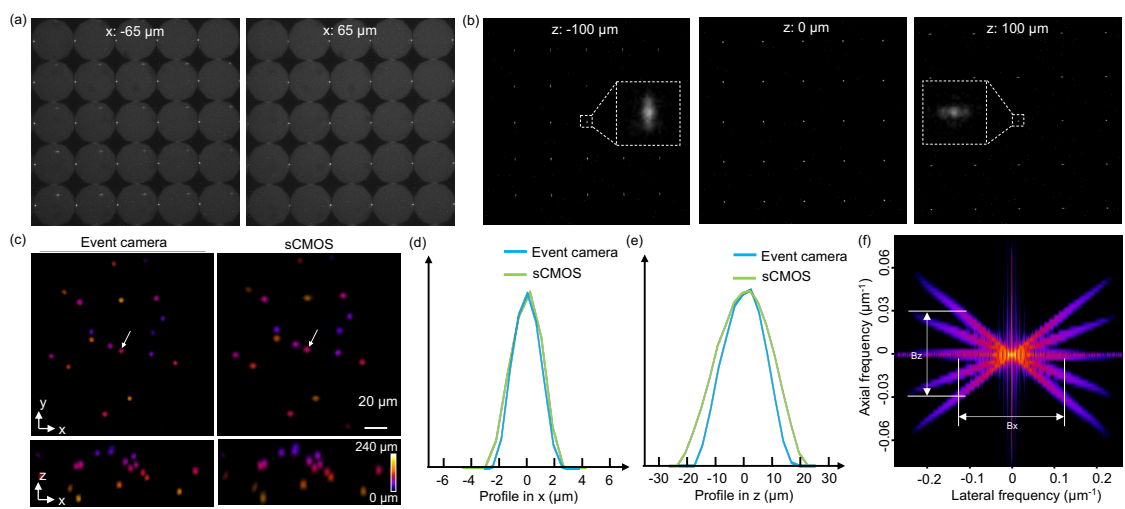

Figure S2: **System characterization**. (a) Two frames captured from a single bead by the traditional Fourier LFM, showcasing the PSFs at leftmost and rightmost boundary of the FOV. (b) Three frames captured at three z positions, including -100 µm, 0 µm and 100 µm, using the same Fourier LFM system. Notable aberrations become evident at the -100 µm and 100 µm axial positions as illustrated in the zoom-in regions. (c) Depth color-coded map of a reconstructed frame from the EventLFM system and Fourier LFM system. (d, e) the intensity profiles along the x and z axis for the bead indicated by the arrow in (c). (f) 3D MTF of the Fourier LFM system.

**Resolution Evaluation**: To evaluate the system's resolution, we utilized a phantom consisting of 2-µm fluorescent beads. We employed the refocusing algorithm to achieve 3D reconstruction, resulting in a color-coded depth map of a reconstructed frame, as shown in Fig. S2(c). Lateral and axial resolution metrics were extracted from the full-width at half-maximum (FWHM) measurements taken along profiles across a selected bead, as displayed in Fig. S2(d) for the x axis and Fig. S2(e) for the z axis. These measurements yielded a FWHM of 3.9 μm in the x dimension and 21.0 μm in the z dimension, findings that are in concordance with the 3D Modulation Transfer Function (MTF) shown in Fig.S2(f).

## 2. Detailed computational pipeline for EventLFM reconstruction

We provide a detailed reconstruction pipeline for EventLFM in Fig. S3. Initially, the raw space-time event stream is processed via a time surface algorithm, effectively converting the dynamic event stream into frames. Following this conversion, a median filter is applied to the time surface frame, aiming to remove the discretized noise. Subsequently, the denoised frame is segmented into 5 x 5 views, with each view representing the viewpoint of an individual microlens. These views are shifted at subpixel resolution and cumulated in a series of refocusing images, each corresponding to a different focal depth. These refocusing images are then sequentially arranged into a volume with a 2 μm depth interval. The culmination of this process involves a thresholding technique, strategically applied to the refocusing volume to eliminate artifacts, thereby ensuring a high-quality reconstruction. Finally, the refocusing volume is then filtered by a threshold to mitigate the refocusing artifacts.

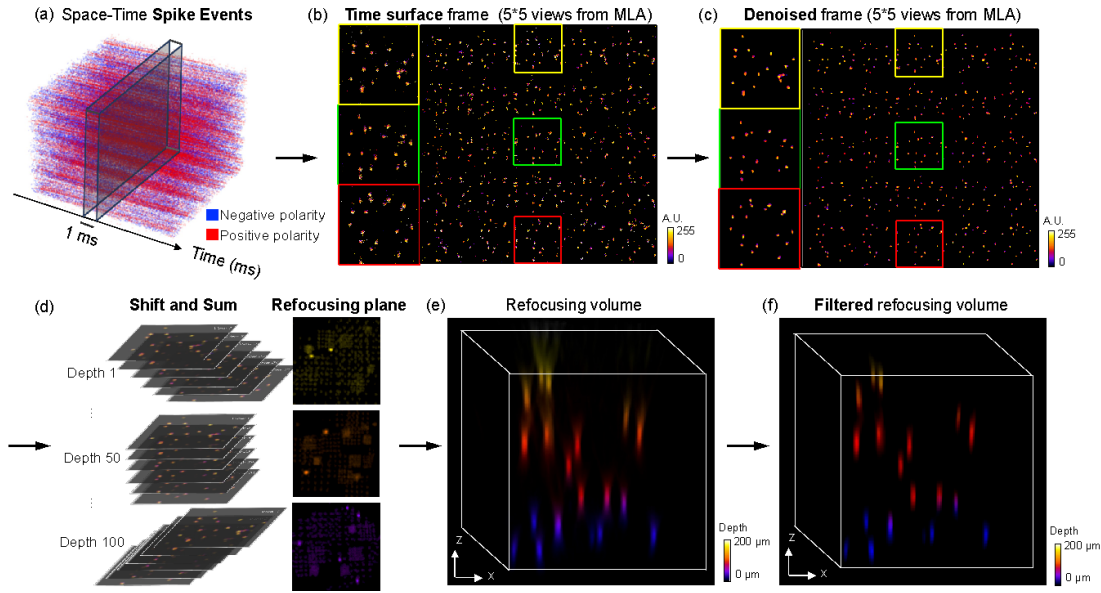

Figure S3: **Detailed computational pipeline**. (a) Space-time event spike stream captured by the event camera, where each event is characterized by its polarity, timestamp, and spatial coordinates. (b) The raw event stream is transformed into a single frame via the time surface algorithm with an integration time of 1 ms. (c) The time surface frame is denoised via a medium filter to enhance the SNR. (d) The denoise frame is cropped into 5x5 views, which are then realigned and cumulatively added (indicated by the orange arrow) to form the refocusing planes at different depths. (e) The refocusing volume is a sequence of 100 refocused planes with a 2 μm depth interval. (f) The refocusing volume is filtered via a threshold to reduce the artifacts.

## 3. Time-surface algorithm

The notion of a "time surface" is a concept used in the domain of event cameras, serving to enrich the representations for recognition tasks[1,2]. In this context, an algorithm is employed to harness an exponential time-decay function, thus generating a time surface that encapsulates both the spatial and temporal correlations existing among adjacent pixels. Subsequently, the pixel value comprising this surface, signifying the signal intensity, are assigned values ranging from 0 to 255, as shown in Fig. S4(a). When a pixel is selected as the most recent event, all other pixels within the temporal stream undergo value assignment contingent upon their respective time-stamps and the decay curve, with earlier events being ascribed smaller values. This approach offers a spatiotemporal context for each specific event, thereby effectively minimizing undesirable motion blurs.

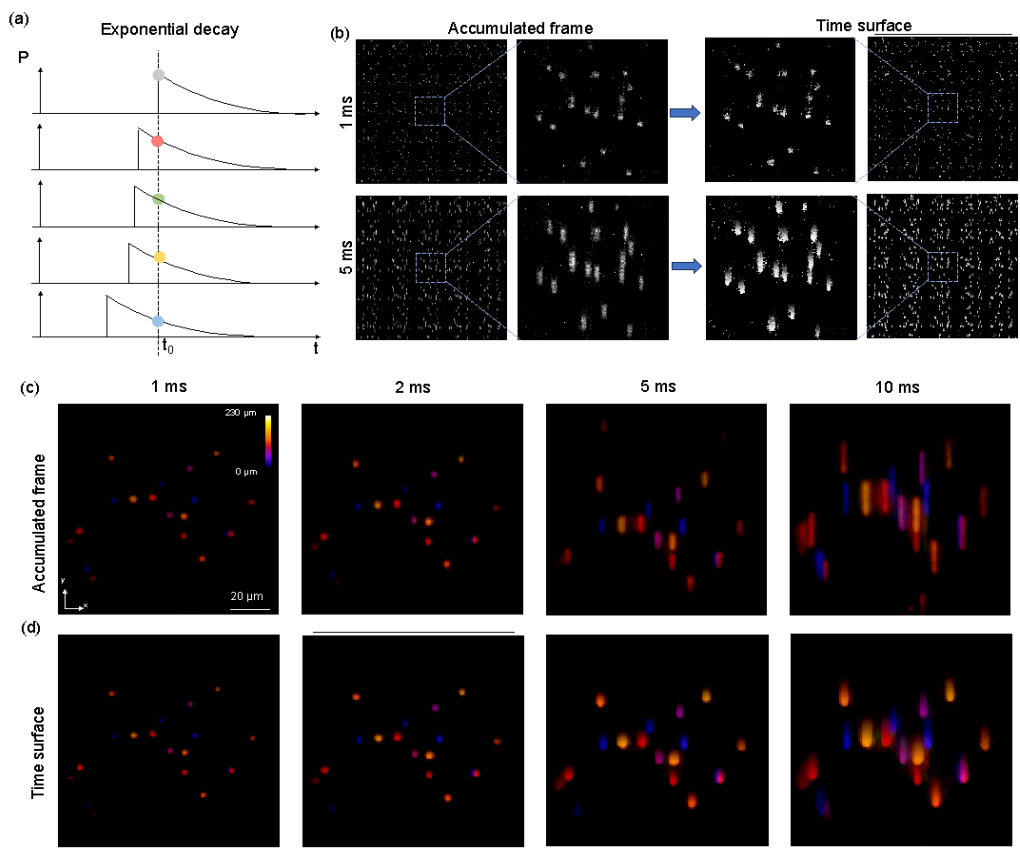

Figure S4: **Time-surface algorithm**. (a) Exponential decay curve used in time-surface algorithm. (b) Visual comparison between directly-accumulated frames and time-surface frames. Top row: 1 ms accumulation time. Bottom row: 5 ms accumulation time. (c) The color-coded depth maps of reconstructed frame generated through the directly-accumulated frames with accumulation times ranging from 1 ms to 10 ms. (d) Color-coded depth maps of reconstructed frames using time-surface frames with accumulation times ranging from 1 ms to 10 ms.

Besides the time-surface algorithm, an alternative method of translating events stream into conventional image frames is the direct accumulation of all events within a defined temporal window into a single frame. A visual comparison of outcomes resulting from the time-surface algorithm and the direct event accumulation approach for a fast-moving object is provided in Figure S4(b). When an accumulation time of 1 ms is employed, both methods yield comparable results. However, when a longer accumulation time is employed, the time-surface method yields frames where motion trails appear less pronounced. Similarly, for dynamic signals with varying velocities processed using specific accumulation times, the time-surface method excels at suppressing motion-induced blurs in the resulting frames. We further perform light field reconstructions utilizing data from both directly accumulated frames and time-surface frames, the resulting volumes represented in color-coded depth maps are shown in Fig. S4(c,d). For this experiment, the speed of the object is set at 2.5 mm/s. When employing a lower accumulation time, both methods yield similar results. However, with an increase in the accumulation time, the efficacy of the time-surface algorithm in mitigating motion blur artifacts (Fig. S4(c)) becomes apparent as compared to the direct-accumulation method (Fig. S4(d)). This demonstrates the superior capability of the time-surface algorithm for processing event streams from dynamic objects.

## *4.* Additional details of denoising process and noise analysis

**Details of the denoising process for biological samples**:

The denoising efficiency of our reconstruction algorithm hinges on two critical operations: the median filter and the summing operation within the refocusing algorithm. Fig. S5 illustrates the impact of these operations on noise reduction during the imaging of neuron-labeled brain slices.

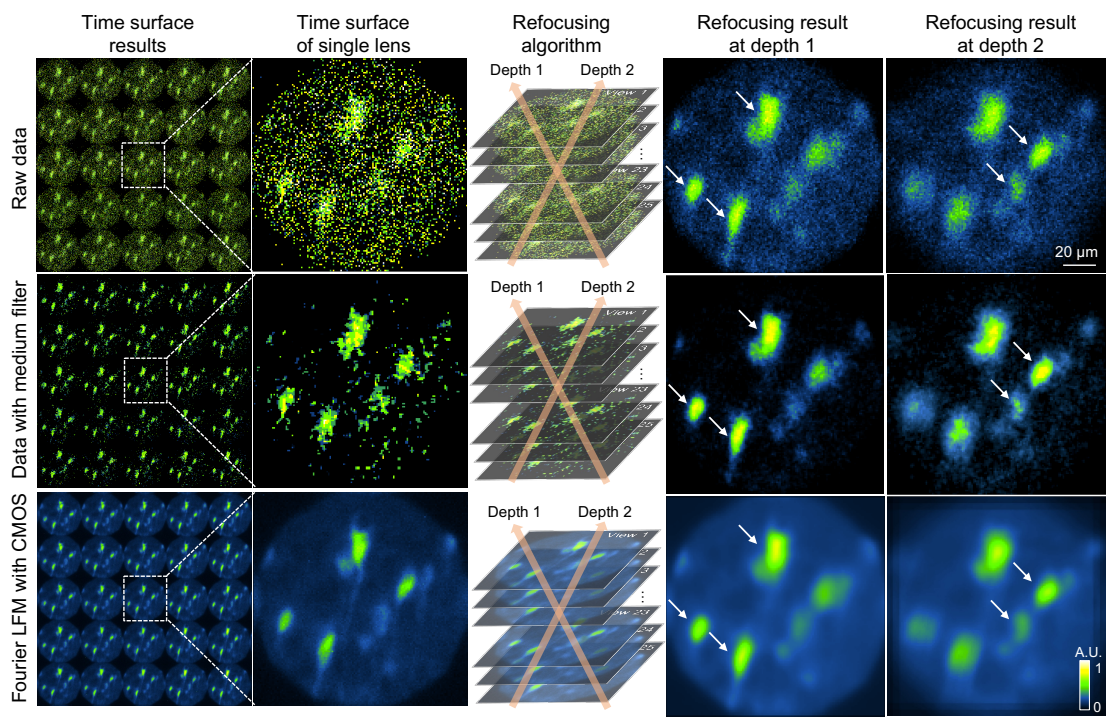

Figure S5: **Detailed denoising process for brain slice**. The first and second rows compare the refocusing results without and with a medium filter. The medium filter effectively removes the discretized noise from the highly sensitive event sensor. Further enhancement of SNR is achieved through the shift-and-sum refocusing algorithm. The refocusing results from EventLFM are validated with traditional Fourier LFM. The white arrows indicate how different neurons are focused on different depths. The comparison with Fourier LFM shows that the EventLFM can accurately reconstruct depth information and suppress the background noise without sacrificing spatial resolution.

Initially, the raw data's time surface is heavily overwhelmed by the sensor noise (shown in the first row of Fig. S5), applying a median filter effectively removes a substantial amount of the noise while preserving the neuronal structures. This efficiency stems from the inherent characteristics of event cameras, which, unlike traditional sensors that capture images at predetermined intervals, asynchronously detect illumination changes at the pixel level. This process naturally leads to a more 'discretized' form of noise, which can be mitigated via medium filtering. Subsequent enhancement of the signal-to-noise ratio (SNR), by approximately 5X, is facilitated by the summing of 25 aligned cropped views within the refocusing algorithm. This enhancement stems from the fact that noise across different views is *uncorrelated*, which permits the summation process to effectively reduce noise while amplifying the signal of in-focus neuronal structures.

Validation of the reconstructed images is achieved through comparison with a conventional Fourier LFM system. The alignment of white arrows in both Fourier LFM and EventLFM, highlighting neurons at various refocusing depths, demonstrates the ability of the reconstruction pipeline to accurately recover neuronal depth information. Furthermore, the event camera's inherent sensitivity to brightness variations provides an intrinsic advantage in suppressing out-of-focus blur and background noise. Consequently, this leads to a cleaner background and higher spatial resolution compared to conventional Fourier LFM.

**Noise analysis for EventLFM with different sensitivity**:

We follow a similar procedure in [4] to characterize the noise characteristics in EventLFM. First, electronic noise is tested when the EventLFM is in a dark environment. Initially, we record the event stream with three different sensitivity thresholds. Then we accumulate all the noise events together and calculate the number of events per pixel and per second as shown in Fig. S6(a). Second, we quantify photon noise by measuring the random noise with a constant illumination onto the event sensor. To quantify the photon noise, a square light source, with an intensity approximating neuronal fluorescence signal, is employed as a constant illuminator. We accumulate the photon noise events within the illuminated square and calculate the number of events per pixel per second as shown in Fig. S6(b). Photon noise dominates over electronic noise at high sensitivity setting, which is the sensitivity setting we used in this study. The noise is mainly isolated pixels randomly distributed across the FOV. We can use a medium filter to mitigate them. In addition, the refocusing algorithm is beneficial in noise suppression.

We evaluate the noise suppression effect of our algorithm by processing the recorded stream from a blinking square uniform source. We calculate the mean as the signal and the standard deviation (std) as the noise, and signal-noise-ratio (SNR) is defined as the ratio between the mean and the std. Fig. S6(c) illustrates the SNR of the raw images, the median-filtered images, and refocused images across three sensitivity settings, corroborating the enhancement of SNR through both the median filter and the refocusing algorithm.

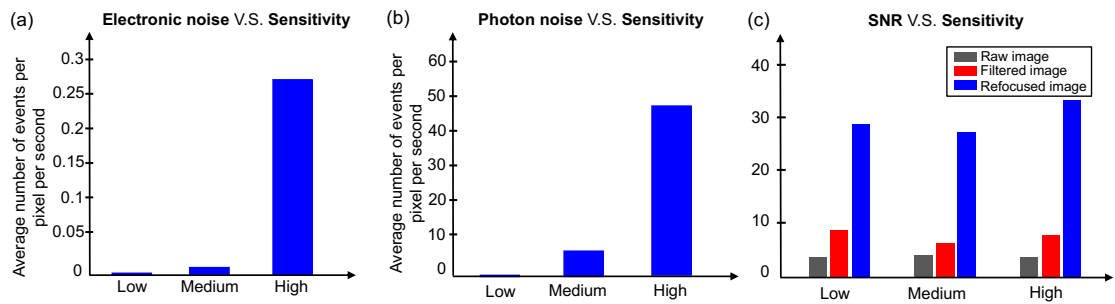

Figure S6: **Noise analysis for EventLFM**. (a) Events induced by electronic noise are obtained in complete darkness. (b) Events induced by photon noise are recorded under constant illumination. (c) SNR comparison across raw image, processed image with medium filter, and refocused image with refocusing algorithm at three different sensitivity settings. Medium filter and refocusing algorithms both improve the SNR.

**Selection of event camera setting:**

The response threshold is a balance between sensitivity and background noise, as shown in Fig. S6. Reducing the threshold improves the pixel sensitivity, and increases pixel response speed, while also inducing more background noise. The selection of an appropriate response depends on the application's demands, with applications requiring a faster response time and detection of subtler brightness changes necessitating a lower threshold. We set it to ensure sufficient sensitivity at practical fluorescence imaging conditions in this work. The integration time directly determines imaging speed. The shorter the integration time is set, the faster the system could achieve. On the other hand, shorter integration time means lower signal. So, we set the response threshold lower to improve the pixel sensitivity and increase pixel response speed, which also induces more background noise and result in low SNR. This highlights the unique challenge in this work that achieves kHz 3D imaging under low SNR situation. The main criterion is to ensure high quality imaging under low SNR.

## 5. Comparison between EventFLM and traditional Fourier LFM for imaging fast-moving objects

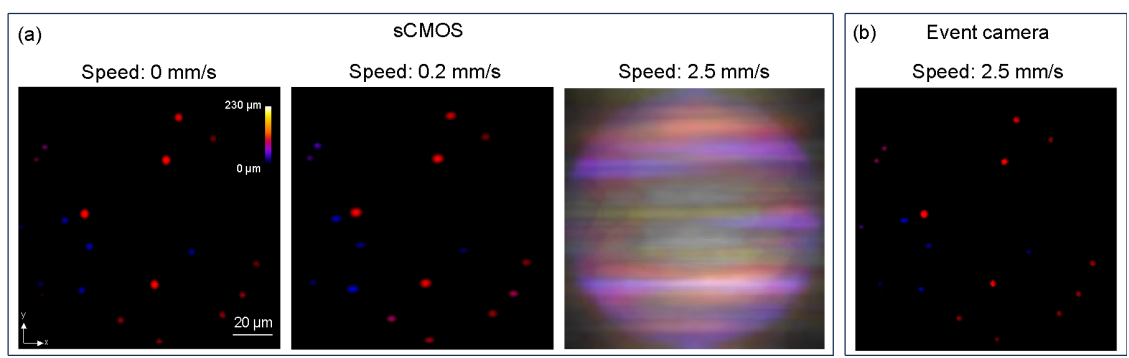

Figure S7: **Comparison between EventLFM and Traditional Fourier LFM for imaging fast moving objects.** (a) Color-coded depth map of reconstructed frames from the same object moving at varied speeds, captured by the traditional Fourier LFM equipped with an sCMOS camera. (b) Color-coded depth map of the reconstructed frame from the same object moving at 2.5 mm/s, acquired with EventLFM.

To elucidate on the superior capability of EventLFM in capturing fast-moving objects, we conduct the following controlled experiments to benchmark the EventLFM's result with the traditional Fourier LFM. First, to demonstrate the performance of the traditional Fourier LFM for imaging moving object, we conduct a series of tests using 3D phantoms moving at various speeds. Initially, the object was imaged while at rest, and reconstruction was performed by employing the standard light field refocusing algorithm, serving as a benchmark. Subsequently, the object's speed was set at 0.2 mm/s, and a sequence of frames was captured for subsequent reconstruction. Following this, the speed was adjusted to 2.5 mm/s, and the same region of the sample was imaged. Upon completion of the reconstruction process, frames containing identical region of interest were selected to generate the color-coded depth map, as depicted in Fig. S7(a). The MIPs obtained when the object was stationary and at a speed of 0.2 mm/s displayed close agreement, validating the consistency of the imaging process. However, when the object's speed increased to 2.5 mm/s, the reconstructed results exhibited considerable motion blur artifacts. The artifacts can be attributed to the constrained frame rate (30 fps) of the sCMOS camera employed. To demonstrate the unique capability of EventLFM for imaging fast-moving object. simulta- neous imaging of the object moving at 2.5 mm/s was conducted with an event camera. Subsequently, reconstruction was performed by employing the same refocusing algorithm, as depicted in Fig. S7(b). Remarkably, the reconstruction closely matched the results obtained with the sCMOS camera under slow-moving conditions.

## 6. Imaging of moving object in z direction

In addition to evaluating the performance of EventLFM for objects moving along the y-axis, we extend our assessment to include objects moving along the z-axis. Fig. 2(f) illustrates the relationship between the accumulation time and the quality of the reconstructed images. An increase in accumulation time results in a decrease in resolution, with the emergence of a tail artifact attributable to the object's high velocity. For this experiment, we utilize a phantom moving along the z-axis at a controlled speed of 0.9 mm/s, a rate dictated by the working distance limitations of our setup. We process the stream with varying accumulation times, ranging from 5 ms to 40 ms. Fig. S6 presents the MIPs of the reconstructed images in both the x-y and y-z planes. Consistent with observations from the phantom moving in the y direction, we note that the apparent size of the reconstructed beads expanded with increasing accumulation time, and a pronounced tail appears at higher accumulation time.

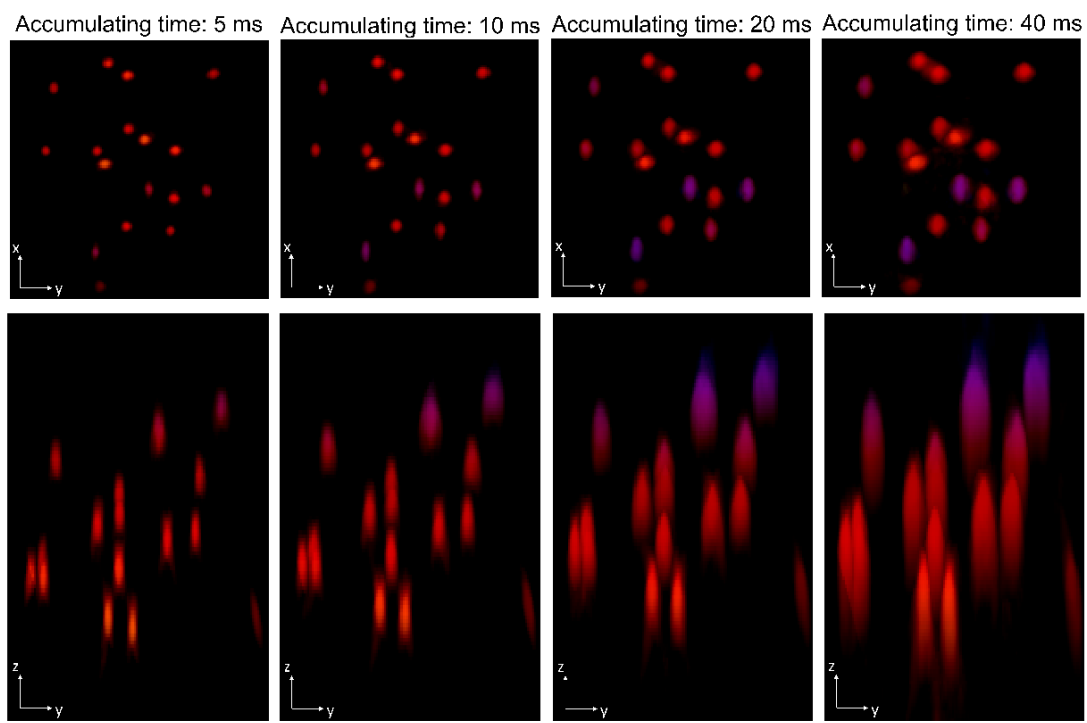

Figure S8: **Reconstructed results of phantom moving in *z* direction.** With the same recorded stream, we use four different accumulation time: 5 ms, 10 ms, 20 ms, and 40 ms. We calculate the MIPs in both x-y plane and y-z plane.

## 7. Comparison between EventFLM and traditional Fourier LFM for imaging neuron labeled *C. elegans*

In order to assess the applicability of our EventLFM system for biological specimens, we conducted imaging of GFP-labeled neurons within multiple freely moving C. elegans. These C. elegans were positioned on a gel substrate and subsequently exposed to a droplet of Basel solution. Within the C. elegans body, four brightly labeled neurons are present, with two located in the tail region and two in the mid-body section. When observed through a fluorescence microscope, these neurons are readily visible. However, by increasing the accumulation time of the event camera, we were able to discern the vague outline of the C. elegans body, as depicted in Fig. S9(a). Dashed lines approximately delineate the positions of the C. elegans within the FOV. Signals from the C. elegans were simultaneously captured using both LFM systems using the event camera and an sCMOS camera, resulting in reconstructions presented in Fig. S9(b). Since the event camera only detects moving or blinking signals. Consequently, the two neurons at the bottom, which remained relatively stationary during recording, are largely absent from the reconstruction obtained from the event camera. For a more detailed examination, we conducted a zoomed-in analysis and traced the trajectories of the four neurons within the central area, as illustrated in Fig. S9(c). This analysis distinctly reveals the movement of the neurons along their respective trajectories within the 3D space.

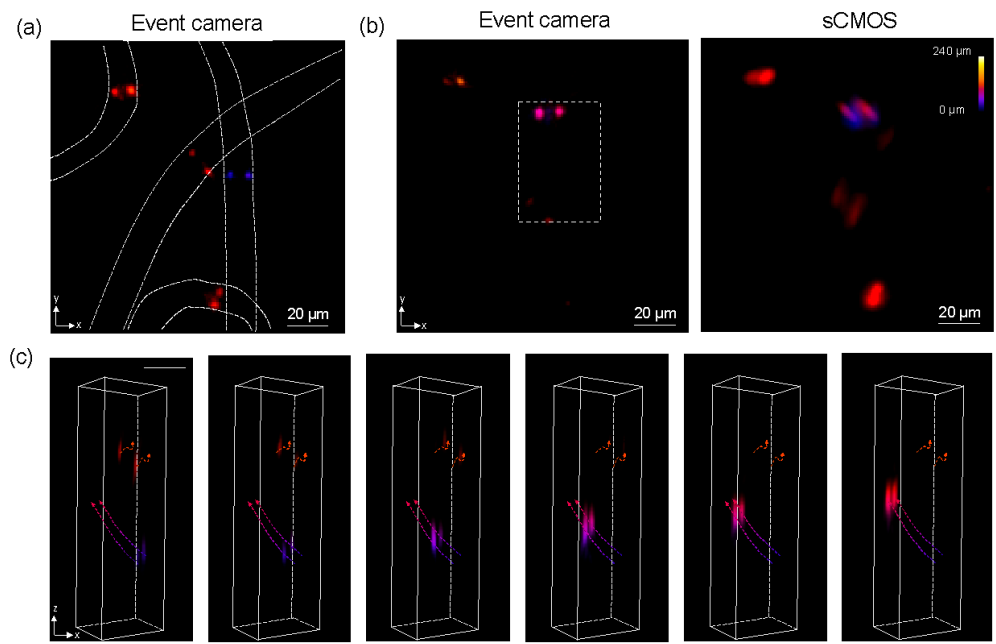

Figure S9: **Additional imaging results of freely-moving *C. elegans*.** (a) An example of a *C. elegans* specimen used in our experiment captured by an event camera. The dashed line delineates the contour of *C. elegans* bodies. (b) Comparison of color-coded depth maps from the reconstructed volumes using EventLFM and the traditional Fourier LFM. (c) 3D zoom-in reconstructions and the tracked trajectories of neurons from EventLFM.

## *8.* **More results from blinking neurons in mouse brain slice**

To demonstrate EventLFM's potential for neural imaging, we image a 75 µm thick section of GFP-labeled mouse brain tissue. The sample is illuminated using a pulsed LED source, designed to simulate neuronal activities within scattering biological tissues. The illumination pulse sequence is set with a 1 ms pulse width and intervals varying from 2 ms to 50 ms. To validate the spatial reconstruction accuracy of EventLFM, we capture the fluorescence signals with traditional Fourier LFM and conventional fluorescent microscopy under constant illumination. Fig. S10(a) shows

MIPs from a single reconstructed frame of each method. By visual inspection, the reconstruction from EventLFM is consistent with Fourier LFM, effectively capturing all neurons within the FOV and the intensity variations among them. However, a notable difference arises in the signal-to-background ratio (SBR). Fourier LFM suffers from a low SBR due to tissue scattering, which results in neuronal signals being buried in strong background fluorescence. In contrast, EventLFM demonstrates a significantly improved SBR, yielding a reconstruction with markedly improved image contrast and suppressed background fluorescence. This improvement is attributed to the event-based measurement mechanism intrinsic to EventLFM, wherein a readout is generated only when intensity changes exceed a certain threshold. Consequently, temporally slowly varying background fluorescence signals, which do not often meet this criterion, are either removed or substantially reduced in the raw data. Additionally, to underline EventLFM's capability of precisely recording fast neuronal spikes, temporal traces from three distinct neurons are extracted, as shown in Fig. S10(b). These traces exhibit a strong correlation with the input illumination pulse sequence, thereby validating that EventLFM can accurately reconstruct neuronal blinking dynamics within scattering tissue.

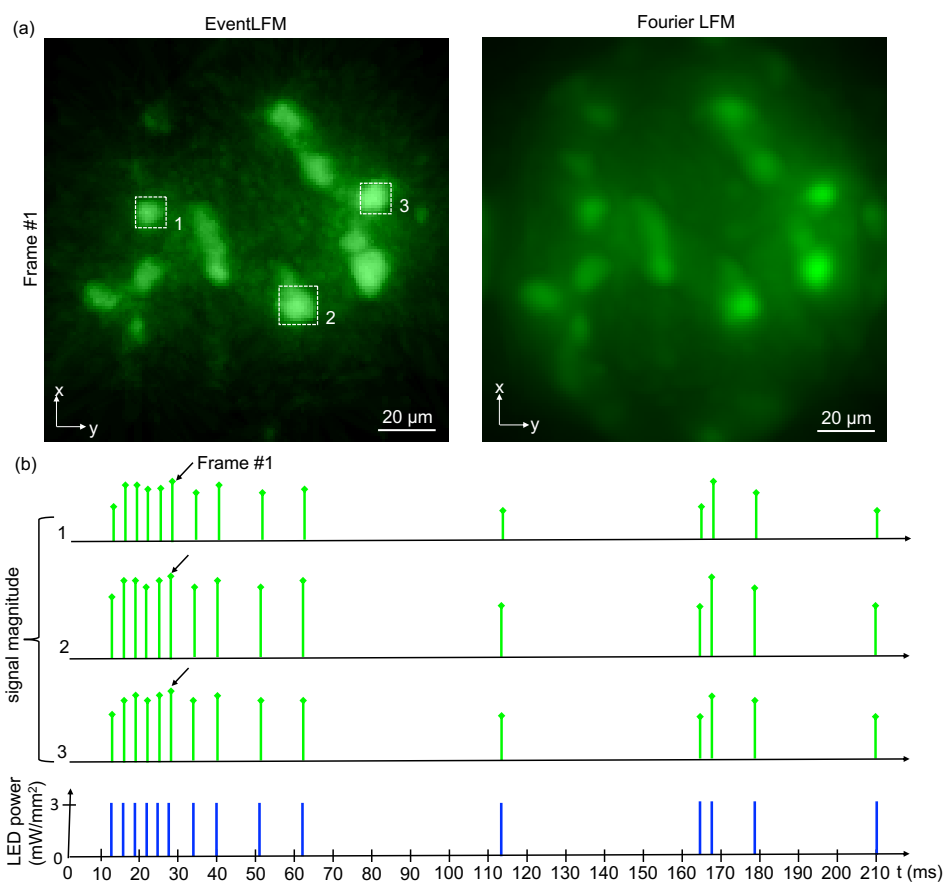

Figure S10: **EventLFM imaging of blinking neurons in mouse brain tissue.** (a) MIPs of single-frame reconstructions, from EventLFM and Fourier LFM, capturing blinking neuronal signals within a 75 µm-thick mouse brain slice emulated with pulsed LED illumination. (b) Temporal traces from three distinct neurons, obtained by calculating the mean intensities in the dashed rectangles labeled 1, 2, and 3 in (a). The LED pulses are set at 1 ms width, with the intervals varying randomly from 2 ms to 50 ms. The reconstructed traces agree with the LED pulse sequence, validating EventLFM's ability to accurately capture blinking dynamics within scattering brain tissue.

## *9.* Additional details on imaging brain slice with target illumination

**Hardware setup**:

A DMD (DLI4130 0.7 XGA VIS High-Speed Kit, Digital Light Innovations) is set up in illumination path to manipulate the distribution of excitation in the FOV. Beam is illuminated onto the surface of the DMD with a degree of 24° through a TIR prism. The DMD surface is imaged onto the FOV with a demagnification of ~40 by a 4F system and the objective. By designing the pattern on the DMD, we can generate spatiotemporal illumination as needed.

**Pattern generation**:

First, we place a fluorescent plate on stage and use the DMD to display a designed pattern. The fluorescence excited by the DMD pattern is captured by the reference camera. Utilizing the *imtransform* function within MATLAB, a transformation matrix is extracted, establishing a correspondence between a designated DMD pattern and its associated image. Then a brain slice replaces the fluorescent plate in the FOV and the reference camera captures the neurons in the FOV under uniform illumination. By employing the transformation matrix, a corresponding DMD pattern is generated, specifically tailored to the neurons of interest within the brain slice. Individual patterns from the neurons are isolated into separate frames. These frames are then integrated into a sequence at varying intervals. Upon activation of the DMD, it projects this sequence of patterns, thereby achieving targeted illumination of the predetermined neurons.

**Additional 3D reconstruction results**:

We provide an additional comparison of the 3D information of the mouse brain tissue detailed in manuscript Section 3.3. Fig. S11 compares the axial profile along the dashed line among axial scanned widefield measurement with uniform and targeted illumination, and EventLFM with targeted illumination. The axial displacement of neurons from EventLFM imaging results are consistent with the widefield system, demonstrating EventLFM's capability to capture 3D information from the brain slice tissue. Moreover, targeted illumination significantly suppresses background noise originating from out-of-focus regions or scattered signals from other neuronal structures, which substantially improves both the SNR and SBR for neuronal activity detection.

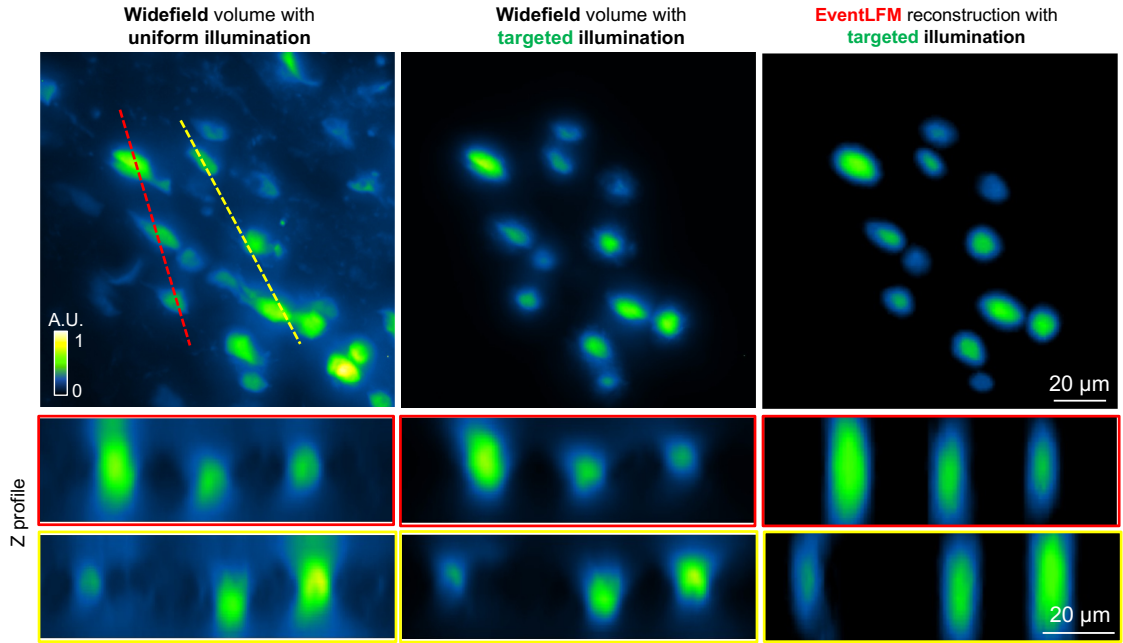

Figure S11: Comparison of the XY MIP and z profile traced along the dashed red and yellow lines across the axial scanned widefield measurement with uniform illumination and targeted illumination, and EventLFM with targeted illumination.

## 10. Convolutional Neural Network for EventLFM reconstruction.

To further demonstrate the potential imaging capabilities of EventLFM, we have integrated a deep learning framework tailored for 3D reconstruction. Preliminary results with fast-moving and dynamic-blinking phantoms embedded with fluorescent particles demonstrate that EventLFM augmented by deep learning facilitates significant improvements in 3D resolution and overall reconstruction quality.

**Network structure**:

Our EventLFM-Net draws inspiration from the CM$^2$Net, which is tailored for high-resolution volumetric reconstruction in imaging systems with multi-view geometry[3]. The network contains two main modules: the view synthesis module and the LFR enhancement module (shown in Fig. S12). The view synthesis module extracts the subpixel parallax information directly from the raw view stack, while the LFR enhancement module enhances upon the initial refocusing volume. Collectively, the network effectively suppresses the reconstruction artifacts from the 'shift-and-add' algorithm and enhances the 3D resolution.

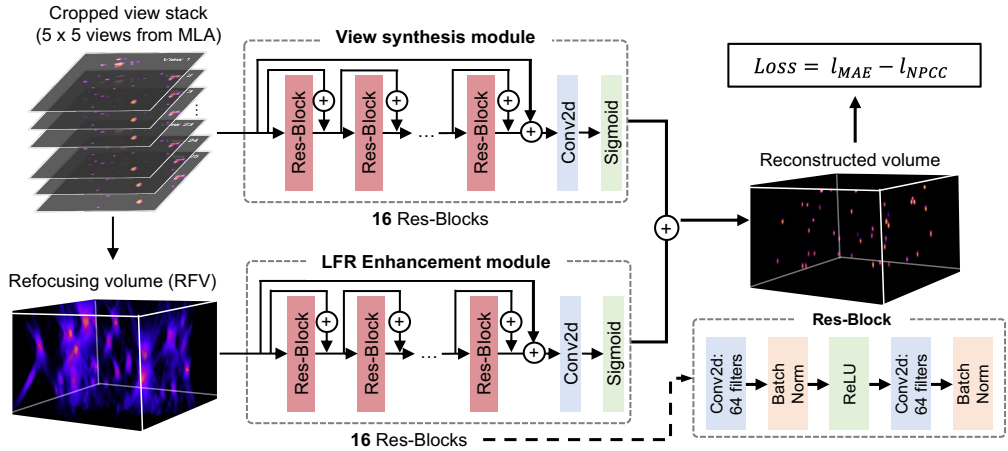

Figure S12: **EventLFM-Net structure**. The input of EventLFM-Net consists of a cropped 5 x 5 view stack alongside the refocusing volume. The view stack is sent to a view synthesis module to integrate the information from multiple lenses. Concurrently, the refocusing volume is sent to an enhancement module to suppress the artifact and improve the 3D resolution from the refocusing volume. The two modules together yield a reconstructed volume with improved uniformity and resolution. The loss is performed by comparing the reconstructed volume with the widefield scanned ground truth.

Both modules have the same backbone structure, containing 16 Res-Blocks with additional skip connections to facilitate multi-scale feature fusion. The loss function is a sum of the Normalized Pearson Correlation Coefficient (NPCC) and Mean Absolute Error (MAE). This dual loss leverages NPCC to refine spatial alignment and MAE to minimize the intensity difference between the 3D reconstruction and ground truth, thus ensuring a comprehensive optimization of both spatial and intensity accuracy. The network is implemented with PyTorch and runs on an Nvidia GPU RTX 4090, with a batch size of 8. The entire training process takes 24 hours.

**Training data collection**: We utilize a phantom with a thickness of 200 μm to build the training dataset. First, we set the speed of the stage as 2.5 mm/s and recorded the stream with EventLFM when the sample is moving in y-direction. Then we move the sample back to the starting position and scan the sample in z direction step by step with the reference conventional wide-field microscopy (20X, 0.75 NA). The depth range is 210 μm with a scanning step size of 3 μm. The sample is moved to the next y position when the z-scanning is done. It takes about 4 hours to scan the whole area of the sample that is recorded by the event camera. We can crop the ground truth according to the refocusing results of the stream from the camera. In total, we collect a training dataset containing 4700 pairs of event stream and wide field ground truth.

**Blinking result on dynamic blinking object**:

The network trained on fast-moving objects is subsequently evaluated with a dynamic blinking object. The experiment setup is detailed in the manuscript Section 3.2. Fig. S13(a) presents the depth-color-coded RFV and CNN reconstruction of frame #1. Consistent with the results from fast-moving objects, the CNN reconstruction of the blinking object also outperforms RFV, showcasing enhanced resolution with improved image quality.

It is noteworthy that the event-based signal characteristics of blinking objects are distinct from those associated with fast-moving objects, as detailed in the manuscript. Despite these differences, by utilizing RFV as a preliminary estimation and conditioned on the network, our trained CNN can generalize across both moving and blinking phantoms. Additionally, Fig. S13(b) illustrates the temporal trace of three particles within the reconstruction, closely matching with the sequence of LED illumination pulses. This analysis confirmed that the network efficiently captures high-frequency blinking signals in a 3D context at enhanced resolutions, demonstrating its versatility and effectiveness in diverse dynamic imaging scenarios.

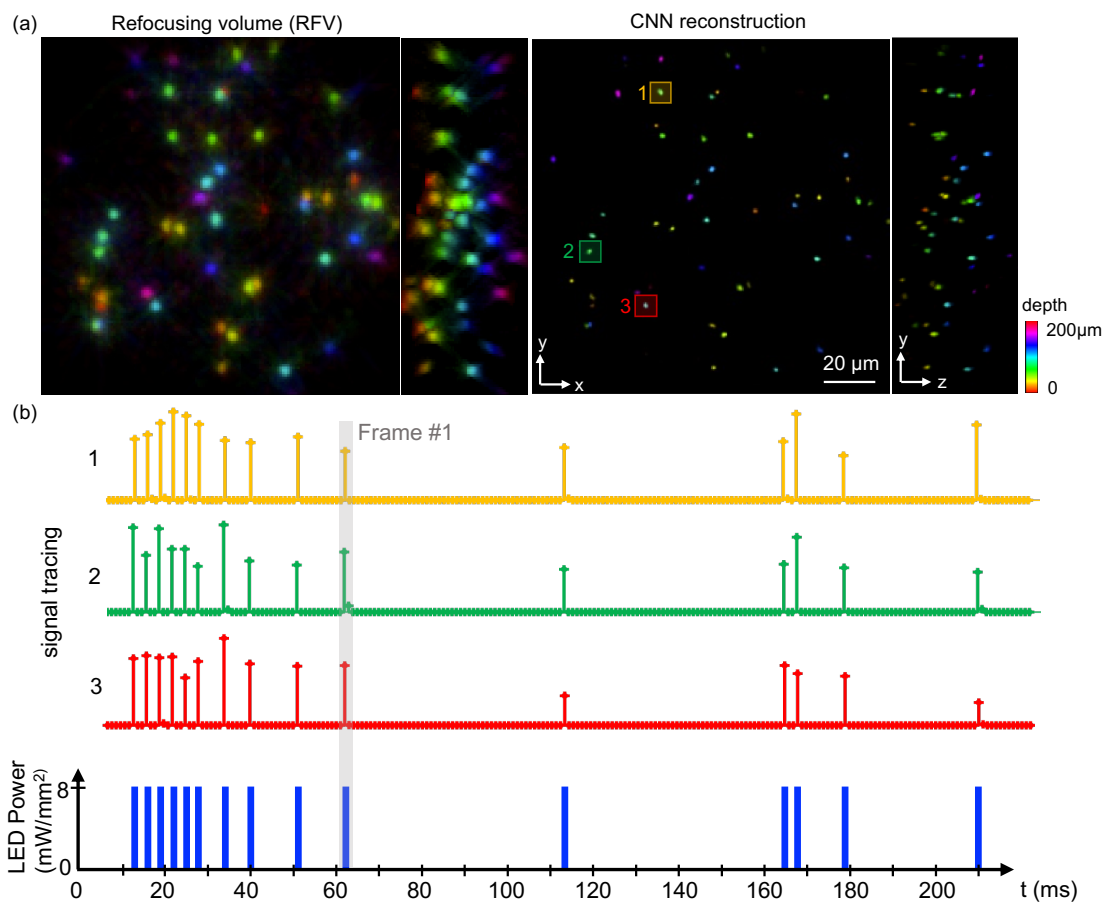

Figure S13: **Imaging results of a dynamic blinking object**. (a) Depth color-coded RFV and 3D CNN reconstruction of a frame marked as Frame #1. The CNN robustly generalizes to dynamic blinking objects, yielding improved image quality and enhanced resolution compared to RFV. (b) Temporal trace analysis for three particles, labeled 1, 2, and 3, within the CNN 3D reconstruction. The LED pulse widths are uniformly set at 1 ms with the inter-pulse intervals randomly varying between 2 ms and 50 ms.

## 11. Visualization 1 (Separate file)

The video shows the motion of beads with different directions and speeds. The color represents the depth information. A transition in color corresponds to bead displacement in the z direction. The frame rate of the video is slowed down to 30 fps for visualization, though it was originally captured at a rate of 1000 fps.

# References

1. Lagorce, X., Orchard, G., Galluppi, F., Shi, B. E. & Benosman, R. B. HOTS: A Hierarchy of Event-Based Time-Surfaces for Pattern Recognition. *IEEE Trans. Pattern Anal. Mach. Intell.* **39**, 1346–1359 (2017).

2. Sironi, A., Brambilla, M., Bourdis, N., Lagorce, X. & Benosman, R. HATS: Histograms of averaged time surfaces for robust event-based object classification. in *Proceedings of the IEEE conference on computer vision and pattern recognition* 1731–1740 (2018).

3. Xue, Y., Yang, Q., Hu, G., Guo, K. & Tian, L. Deep-learning-augmented computational miniature mesoscope. *Optica* **9**, 1009 (2022).

4. C. Cabriel, T. Monfort, C. G. Specht, and I. Izeddin, "Event-based vision sensor for fast and dense single-molecule localization microscopy," *Nat. Photonics* **17**, 1105–1113 (2023).